\documentclass[11pt,a4paper]{article}
\usepackage{ifpdf}
\usepackage{ae}
\usepackage[ansinew]{inputenc}
\usepackage[T1]{fontenc}
\usepackage{mathrsfs}
\usepackage{amsmath}
\usepackage{amssymb}
\usepackage{subcaption}
\pdfoutput=1
\usepackage{amsmath,amssymb,amsfonts,a4wide,graphicx,bm,times,psfrag,wrapfig,sidecap}
\usepackage{cite}
\usepackage[colorlinks=true,linkcolor=black, citecolor=black,
urlcolor=black]{hyperref} 
\numberwithin{equation}{section}
\makeatletter \let\old@startsection=\@startsection
\renewcommand{\@startsection}[6]
{\old@startsection{#1}{#2}{#3}{#4}{#5}{#6\mathversion{bold}}}
\makeatother

\newcommand\re[1]{({\ref{#1}})}
\def\be{\begin{eqnarray}  }
    \def\ee{\end{eqnarray}}

    \def\no{\nonumber}
    \def\la{\label}

    \def\({\left(} \def\){\right)} \def\<{\langle\,} 
    \def\>{\,  \rangle} \def\[{\left[} \def\]{\right]} \def\tr{{\rm tr} }

            \def\CF{{\cal F}}
     \def\p{\partial}  
    \def\g{\gamma}   
    \def\vp{\varphi} \def\th{\theta}  
  
 \def\Tr{{\rm Tr}}

\def\d{\delta}

\usepackage[ansinew]{inputenc}
\newcommand{\imineq}[2]{\vcenter{\hbox{\includegraphics[height=#2ex]{#1}}}}

\begin{document}

\newcommand\encadremath[1]{\vbox{\hrule\hbox{\vrule\kern8pt
\vbox{\kern8pt \hbox{$\displaystyle #1$}\kern8pt}
\kern8pt\vrule}\hrule}} \def\enca#1{\vbox{\hrule\hbox{
\vrule\kern8pt\vbox{\kern8pt \hbox{$\displaystyle #1$} \kern8pt}
\kern8pt\vrule}\hrule}}

\def\Long{{\color{red} $ \Rightarrow$  Long}
\bigskip
}
\def\red{ \color{red} }
\def\blue{ \color{blue}}


\thispagestyle{empty}

\begin{flushright}
\end{flushright}

\vspace{1cm}
\setcounter{footnote}{0}

\begin{center}

 {\Large\bf  Boundary TBA, trees and loops }

\vspace{20mm} 

Ivan Kostov,
 Didina Serban and
Dinh-Long Vu \\[7mm]
 
{\it  Institut de Physique Th\'eorique, CNRS-UMR 3681-INP,
 C.E.A.-Saclay, 
 \\
 F-91191 Gif-sur-Yvette, France} 
 \\[5mm]

\end{center}

\vskip9mm

\vskip18mm


  { \noindent{ 
  We derive a graph expansion for the thermal partition function of  solvable
    two-dimensional  models with boundaries. This expansion of the integration measure over the virtual particles 
     winding around the time cycle is obtained with the help of 
    the matrix-tree theorem.  The free energy is a sum over all connected graphs, 
     which can be either trees or trees  with one loop.  The generating function for the connected 
     trees satisfies a non-linear integral equation, which is equivalent to the TBA equation.
     The sum over connected graphs gives the bulk  free energy as well as the exact $g$-functions
     for the two boundaries.
     We reproduced the integral formula  conjectured by  Dorey, Fioravanti, Rim and Tateo,
     and proved subsequently by Pozsgay. 
Our method can be extended to the case of non-diagonal bulk
scattering and diagonal reflection matrices with proper regularization.
     } }
\newpage
\setcounter{footnote}{0}

\section{Introduction}
\label{sec:1}

The notion of integrability has been extended to systems with
boundary by Ghoshal and Zamolodchikov \cite{Ghoshal:1993tm}.  With the
Yang-Baxter equation, unitarity, analyticity and crossing symmetry for
both bulk scattering matrix and boundary reflection matrix, a model
with integrable boundary is expected to be exactly solved.  One of the
simplest observables in such system is its free energy in large volume limit
and at finite temperature.  Unlike in periodic systems, this free
energy contains a volume-independent correction, also known as the boundary entropy or $g$-function \cite{Affleck:1991tk}.

The first attempt to compute $g$-function was carried out  by  LeClair,  Mussardo,  Saleur and  Skorik \cite{LeClair:1995uf}, using the Thermodynamics Bethe Ansatz (TBA) saddle point approximation.  They  obtained  an expression similar to the  bulk TBA free energy
\begin{align}
\log(g_ag_b)^\textnormal{saddle}(R)=\frac{1}{2}\int_{-\infty}^{+\infty}
\frac{du}{2\pi}\, \Theta_{ab}(u)\log[1+e^{-\epsilon(u)}],\label{TBA}
\end{align}
where $u$ is the rapidity variable, $\epsilon$ is the pseudo-energy at inverse temperature $R$ and $\Theta$ term involves the bulk scattering and the boundary reflection matrices.  

It was later shown by Woynarovich \cite{Woynarovich:2004gc} that another volume-independent contribution is  produced  by  the fluctuation around the TBA saddle point.  The result can be written as a Fredholm determinant
\begin{align}
\log(g_ag_b)^\textnormal{fluc}(R)=-\log\det(1-\hat{K}^+),\label{fluc}
\end{align}
where the kernel $\hat{K}^+$ involves the pseudo-energy and
the bulk scattering matrix but not the reflection matrices.  In other
words, this fluctuation around the saddle point is
boundary independent.  A major problem of Woynarovich's
computation  is that it also predicts a similar term  for periodic
systems, while it is known that  there is no such correction.

Dorey, Fioravanti,  Rim  and   Tateo \cite{Dorey:2004xk} took a different approach towards this problem.  They started with the definition of the partition function as a thermal sum over a complete set of states labelled by mode numbers.  In the infinite volume limit, this sum can be replaced by
integrals over rapidity.  The integrands were explicitly worked
out for small number of particles.  Based on these first terms and the
structure of TBA saddle point result \eqref{TBA}, the authors advanced a conjecture about the boundary-independent part of $g$-function.  Their
proposal has the structure of a Leclair-Mussardo type series
\begin{align}
\log(g_ag_b) (R)&=\log(g_ag_b)^\textnormal{saddle}(R)\nonumber\\
&+\sum_{n\geq
1}\frac{1}{n}\prod_{j=1}^n\int_{-\infty}^{+\infty}\frac{du_j}{2\pi}
\frac{1}{1+e^{\epsilon(u_j)}}K(u_1+u_2)K(u_2-u_3)...K(u_n-u_1),
\label{cluster-expansion}
\end{align}
where $K$ is the logarithmic derivative of the bulk scattering matrix.

Pozsgay \cite{Pozsgay:2010tv} (see also Woynarovich \cite{Woynarovich:2010wt}) argued that the same expression for $g$-function could be obtained from a refined version of TBA saddle point approximation. He noticed that the mismatch between \eqref{fluc} and the series in \eqref{cluster-expansion} is resolved if one uses a non-flat measure for the TBA functional integration.   
This  non-trivial measure comes from  the Jacobian  of the change of variables from mode number to rapidity,  and represents the continuum limit of the   Gaudin determinant.
   
The fluctuation around the saddle point involves only diagonal elements  of this Gaudin matrix, resulting in   the inverse power of  the Fredholm determinant $\det(1-\hat{K}^+)$.  On the other hand, the functional integration measure contains the off-diagonal elements as well, which constitute another Fredholm determinant $\det(1-\hat{K}^-)$.  Pozsgay rewrote the result \eqref{cluster-expansion} in terms of these two Fredholm determinants
\begin{align}
\log(g_ag_b)(R)=\log(g_ag_b)^\textnormal{saddle}(R)+\log\det\frac{1-\hat{K}^-}{1-\hat{K}^+}.
\label{pozsgay}
\end{align}
The two kernels $\hat{K}^\pm$ can be read off from the asymptotic
Bethe equations.  For a periodic system, they happen to be the same
and the effects from the fluctuation and the measure cancel each
other.

It is important to distinguish the Jacobians in \cite{Dorey:2004xk}
from the one in \cite{Pozsgay:2010tv}.  The former appear in each term
of the cluster expansion while the latter is obtained from the
thermodynamics state that minimizes the TBA functional action.  Put it
simply, the Jacobian in \cite{Pozsgay:2010tv} is the thermal average
of all the Jacobians in \cite{Dorey:2004xk}.

In this paper, we derive this known result for $g$-function,
following the strategy of \cite{Dorey:2004xk}: writing the partition
function as a sum over mode numbers and replacing it by an integral over
phase space in the infinite volume limit.  In contrast to
\cite{Dorey:2004xk}, we are able to exactly carry out the cluster
expansion, by virtue of the matrix-tree theorem
\cite{Chaiken82acombinatorial}.  This theorem allows us to write the
Jacobian for a finite number of particles as a sum over diagrams.
Consequently,  the $g$-function is expressed as a sum over  graphs with no loops (trees)  and graphs with one loop.  These combinatorial objects possess simple structure and their sum can be written in the form \eqref{cluster-expansion} or \eqref{pozsgay}.  Compared to \cite{Pozsgay:2010tv}, the Gaussian fluctuations and the measure can be respectively interpreted as the sum over two types of loops. Our final result coincides with the one of \cite{Pozsgay:2010tv}, but our  method  allows an exact treatment of each term in the  canonical partition sum, before the thermodynamical limit. Such advantage makes it potentially useful in the computation of more subtle objects such as correlation functions. This method has been applied for the free energy and one point function of a local observable of a periodic system \cite{Kostov:2018ckg},\cite{article},\cite{PhysRevE.63.036106}. The same idea has been used to derive the equation of state in Generalized Hydrodynamics \cite{10.21468/SciPostPhys.6.2.023}.

Generalization to a theory with $n$ species of particles interacting via diagonal bulk scattering  and diagonal reflection matrices is straightforward.  The graphs now involve $n$ types of vertices and the Fredholm  kernels  are $n\times n$ matrices. We also comment on how the g-function of a theory solved by Nested Bethe Ansatz can be obtained through a regularization procedure. We implement such procedure for a concrete example in another work \cite{Vu2019}.

The paper is structured  as follows. In section 2 we recall 
the  definition of $g$-function and spell out the  Fredholm determinant formula  \re{pozsgay} for a massive theory with diagonal bulk and boundary scattering. In section 3 we develop the combinatorics  needed to sum up  the cluster expansion  and  express the partition function on a cylinder  as a sum over (multi)wrapping virtual particles. In section 4 we   expand, with  help of the matrix-tree theorem, the canonical partition function  on a cylinder as a sum over certain  set of Feynman graphs. In section 5
we perform the sum and recover the expression for the $g$-function. We compute the excited state g-function at the end of the section. In  section 6 we generalise our method to theories of more than one type of particle with diagonal scatterings. We also establish a protocol to deal with g-function of theories solved by Nested Bethe Ansatz. The two appendices present two different proofs of the matrix-tree theorem in the form used in this paper.  
\section{ Bulk and boundary  free energy  of a massive integrable field theory}
\label{sec:2}
The $g$-function, also known as boundary entropy or  
ground-state degeneracy,  was first introduced by Affleck
and Ludwig \cite{Affleck:1991tk}  and since then has been given many physical
interpretations.  In this paper we shall look at this multifaceted
object as the non-extensive contribution to free energy of a system
with boundaries.

Let us  consider an $1+1$ dimensional field theory with a single massive
excitation above the vacuum,  defined  in  an open interval of   length
$L$, whose boundaries  will be denoted by  $a$ and $b$ .  The momentum and
energy of a particle are parameterized by its rapidity $p=p(u), E=E(u)$.
The theory is integrable with a two-to-two bulk scattering phase
$S(u,v)$ and reflection factors $R_a(u),R_b(u)$ at the boundaries. They satisfy a set of conditions \cite{Ghoshal:1993tm}, among which  the unitarity condition
\begin{align}
S(u,v)S(v,u)=R_a(u)R_a(-u)=R_b(u)R_b(-u)=1.\label{unitarity}
\end{align}
The bulk scattering phase does not necessarily depend on the difference between rapidities. We  assume a milder condition
\begin{align}
S(u,-v)S(-u,v)=1,
\label{crossing}
\end{align}
as well as   $S(u,u)=-1$.

The partition   function   at inverse temperature $R$ is defined by the thermal trace
\begin{align}
Z_{ab}(R,L)=\Tr\;e^{-H_{ab}(L)R},
\label{R-channel}
\end{align}
where $H_{ab}(L)$ is the Hamiltonian of the theory living on a segment 
of length $L$  with boundary conditions $a$ and $b$.
One can consider in parallel the partition function of a theory defined on a circle of length $L$
\begin{align}
\la{Zperiodic}
Z(R,L)=\Tr\;e^{-H(L)R}.
\end{align}
The  boundary entropy of the open system is  given by the difference in the two free energies
\begin{align}
\mathcal{F}_{ab}(R,L)\equiv\log Z_{ab}(R,L)-\log Z(R,L).
\end{align}
The $g$-function is defined as the contribution of a
\textit{single} boundary to the free energy.  To compute  it, we pull
the two boundaries far away from each other to avoid interference
\begin{align}
\log g_a(R)=\frac{1}{2}\lim_{L\to\infty}\mathcal{F}_{aa}(R,L).\label{$g$-function-definition}
\end{align}
Compared to the usual definition of $g$-function given in perturbed CFTs
literature, our definition seems to be over-simplifying.  This is due
to our specific choice of normalization of partition functions.  More
precisely, we have fixed the ground state energy in the $L\to\infty$
limit of both Hamiltonians
$H({L})$ and $H_{ab}(L)$ to zero by discarding the bulk energy density
as well as its non-extensive boundary contributions.
\begin{figure}[h]
\centering
\includegraphics[width=10cm]{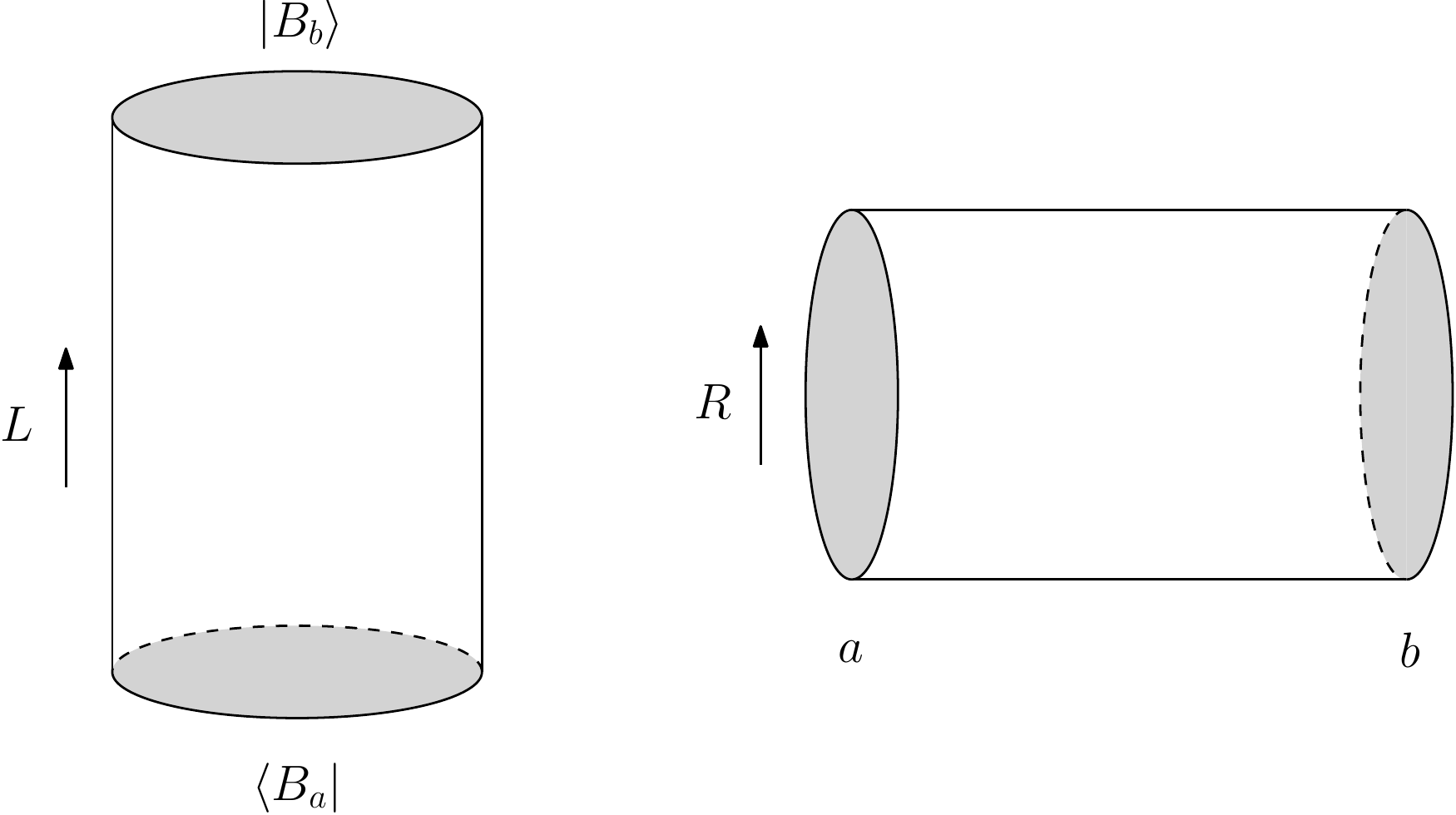}
  \caption{Two equivalent ways of evaluating the partition function on a cylinder.}
\end{figure}

In a relativistic theory there is a mirror transformation
 exchanging the roles of space and time
\be
\tilde p(u) = - i E(u^\g), \ \ \ \tilde E(u) =  - i p(u^\g),
\ee
where  $u^\g$ means analytical continuation  in the rapidity variable which assures that  
the mirror particle has positive energy $\tilde E$ and real momentum $\tilde p$.
 The inverse is true only if the mirror theory coincides with the original one.
 In this case  the natural parametrisation is $p= m\sinh u, E= m\cosh u$ and $u^\g= u+ i\pi/2$. The product of two mirror transformations, $u\to u+ i\pi$,  gives a crossing transformation.

 In terms of the  mirror theory, defined on a circle with circumference $R$, 
 the partition function with periodic boundary conditions \re{Zperiodic}
 takes a similar form
\begin{align}
Z(R,L)=\Tr\;e^{-\tilde{H}(R)L},
\end{align}
 where the trace is in the Hilbert space  of the mirror theory.
 In contrast,  after a mirror transformation the thermal partition function with open boundary conditions   becomes  the overlap of    an initial state
$\langle B_a|$ and a final state  $|B_b\rangle$    defined on a circle 
of circumference $R$ after evolution 
  at mirror   time  $L$    \cite{Ghoshal:1993tm}. 
 Evaluated in the mirror theory, the partition function \re{R-channel} reads
\begin{align}
Z_{ab}(R,L)=\langle B_a|e^{-\tilde{H}(R)L}|B_b\rangle.\label{L-channel}
\end{align}
Although the partition function is the same, the physics is rather
different in the two channels.  In the mirror theory, the $g$-function provides information about
overlapping of the boundary states and the ground state at finite volume.  To see this, we write \eqref{L-channel} as a
sum over eigenstates $|\psi\rangle$ of the periodic Hamiltonian
$\tilde{H}(R)$
\begin{align*}
\langle B_a|e^{-\tilde{H}(R)L}|B_b\rangle=\sum_{|\psi\rangle}\frac{\langle B_a|\psi\rangle}{\sqrt{\langle \psi|\psi\rangle}}e^{-L\tilde{E}(|\psi\rangle)}\frac{\langle \psi|B_b\rangle}{\sqrt{\langle \psi|\psi\rangle}}.
\end{align*}
In the large $L$ limit, this sum is dominated by a single term corresponding to the ground state $|\psi_0\rangle$. The $g$-function is then given by the overlap between this state and the boundary state
\begin{align}
g_a(R)=\frac{\langle B_a|\psi_0\rangle}{\sqrt{\langle \psi_0|\psi_0\rangle}}.\label{overlap-ground}
\end{align}
An expression for $g$-function was conjectured in \cite{Dorey:2004xk}
and proven in \cite{Pozsgay:2010tv}.  Here we  write down  this result  for the case where the  bulk scattering
matrix   is not of difference form. 
Let us denote respectively by $K,K_a$ and $K_b$ the logarithmic
derivatives  of the bulk scattering phase and the boundary reflection
factors associated with the boundaries $a$ and $b$
\begin{align*}
K(u,v)=-i\partial_u \log S(u,v),\quad K_a(u)=-i\partial_u\log
R_a(u),\quad K_b(u)=-i\partial_u\log R_b(u).
\end{align*}
It follows from \eqref{unitarity} and \eqref{crossing} that
\begin{align}
K_a(u)=K_a(-u),\quad K_b(u)=K_b(-u),\quad K(u,-v)=K(-u,v).
\end{align}
Let us also define
\begin{gather}
\Theta_\star(u)\equiv K_\star(u)-K(u,-u)-\pi\delta(u),\quad
\star=a,b\label{Omega-term}.
\end{gather}
Then the expression for $g$-function 
found in \cite{Pozsgay:2010tv}  reads
\begin{align}
\log g_a(R)=\frac{1}{2}\int_{-\infty}^{+\infty}\frac{du}{2\pi}
\, \Theta_{a}(u)\log(1+e^{-\epsilon(u)})
+\frac{1}{2}\log\det\frac{1-\hat{K}^-}{1-\hat{K}^+},
\label{$g$-function}
\end{align}
where $\epsilon$ is the pseudo-energy  at inverse
temperature $R$
\begin{align}
e^{-\epsilon(u)}=e^{-E(u)R}\exp\bigg[\int_{-\infty}^\infty
\frac{dv}{2\pi}K(v,u)\log(1+e^{-\epsilon(v)})\bigg].
\end{align}
The kernels $\hat{K}^\pm$ have support on the positive real axis and
their action is given by
\begin{align}
\hat{K}^\pm F(u)=\int_0^\infty\frac{dv}{2\pi}\big[K(u,v)\pm
K(u,-v)\big]\frac{1}{1+e^{\epsilon(v)}}F(v).
\label{Q}
\end{align}
 In the next sections  we will derive  the expression \eqref{$g$-function}  by evaluating the partition function in the
$R$-channel, namely equation \eqref{R-channel}, in the limit when $L$ is large.  In order to do that,
we will insert a decomposition of the identity in  a complete basis of 
eigenstates of the Hamiltonian $H_{ab}(L)$ and
perform the thermal trace.

\section{Partition function  on a cylinder as a sum over wrapping particles}
\label{section:3}

\subsection{Asymptotic Bethe equations in presence of boundaries}

The $g$-function \eqref{$g$-function-definition} is extracted by taking the limit of  large  volume $L$.
In this limit, we can diagonalize the Hamiltonian $H_{ab}(L)$ using
the technique of Bethe ansatz.

Consider an $N$-particle eigenstate 
$|\textbf{u}\rangle=|u_1,u_2,...,u_N\rangle$.  To obtain the Bethe Ansatz equations
in presence of two boundaries, we follow a particle of
rapidity $u_j$ as it propagates to a boundary and is reflected to the
opposite direction.  It continues to the other boundary, being
reflected for a second time and finally comes back to its initial
position, finishing a trajectory of length $2L$.  During its
propagation, it scatters with the rest of the particles twice, 
once from the left and once from the right.
This process translates into the quantisation
condition of the state $|\textbf{u}\rangle $
\begin{align}
e^{2ip(u_j)L}R_a(u_j)R_b(u_j)\prod_{k\neq
j}^NS(u_j,u_k)S(u_j,-u_k)=1,\quad \forall
j=1,...,N.\label{propagation}
\end{align}
We can write these equations in logarithmic form by introducing a new
set of variables: the total scattering phases
$\phi_1,\phi_2,...,\phi_n$ defined by
\begin{align}
\phi_j(\textbf{u})\equiv 2p(u_j)L-i\log [R_aR_b(u_j)\prod_{k\neq j}^N
S(u_j,u_k)S(u_j,-u_k)],\quad \forall j=1,...,N.\label{Bethe-equations}
\end{align}
In terms of these variables, the quantization of state
$|\textbf{u}\rangle$ reads
\begin{align}
\phi_j(\textbf{u})=2\pi n_j\quad \forall j=1,...,N\quad
\textnormal{with } n_j\in\mathbb{Z}.\label{quantization}
\end{align}
The next step is to impose particle statistics and indistinguishability.
In periodic systems, simply taking the mode numbers $n_j$ to be
all  different automatically satisfies both principles.   However
in the presence of boundary two particles having the same mode numbers
but of opposite signs are indistinguishable.  To avoid 
overcounting of states, we should put a positivity constraint on mode
numbers.


A basis in the $N$-particle sector of the Hilbert space is then labeled by all 
sets of strictly increasing positive integers
 $0<n_1<\dots<n_N$.  The
corresponding eigenvector of the Hamiltonian $H_{ab}(L)$ is characterised by a set of rapidities
$0<u_1<\dots <u_N$, obtained
by solving the Bethe equations \eqref{Bethe-equations} and
\eqref{quantization}.


Inserting a complete set of eigenstates we write the partition function on a cylinder as
\begin{align}
Z_{ab}(R,L)=\sum_{N=0}^\infty\;\sum_{0<n_1<...<n_N}e^{-RE(n_1,...,n_N)}.
\label{sum-constraint}
\end{align}
In this equation, the energy $E$ is 
a  function of mode
numbers $n_1,...,n_N$.   To find its explicit form, one needs to solve the Bethe equations for the
corresponding rapidities $u_1,...,u_N$.  As a function of the rapidities, the energy is 
equal to the sum of the energies of the individual particles
\begin{align*}
E(u_1,...,u_N)=\sum_{j=1}^NE(u_j).
\end{align*}

In order to write the sum \eqref{sum-constraint} as an integral over
rapidities, we first have to remove the constraint between the mode
numbers.  We do this by inserting Kronecker symbols to get rid of
unwanted configurations
\begin{align}
Z_{ab}(R,L)=\sum_{N=0}^\infty\frac{1}{N!}\;\sum_{0\leq n_1,...,n_N}\,\prod_{j<k}^N(1-\delta_{n_j,n_k})\prod_{j=1}^N(1-\delta_{n_j,0})e^{-RE(n_1,...,n_N)}.\label{sum-unconstraint}
\end{align}
%
The first Kronecker symbol introduces the condition that the mode numbers are all different,
 and the second one eliminates the mode numbers equal to zero.

Let us expand  in monomials  the first  factor containing Kronecker symbols, 
which imposes the exclusion principle.  
The partition function \eqref{sum-unconstraint} can be written as a
sum over all sequences   
$(n_1^{r_1},...,n_m^{r_m})$ of non-negative, but otherwise unrestricted 
mode numbers $n_i$ with multiplicities $r_i$.
Each sequence $(n_1^{r_1},...,n_m^{r_m})$ in the sum corresponds to a
state with $r_j$ particles of the same mode number $n_j$, for
$j=1,2,...,m$. The total number of particles in such a sequence is $N=r_1+\dots + r_m$.

For instance, there are four sequences all correspond
to  unphysical state with three particles of the same mode number
$n$: $(n^3),(n^2,n^1),(n^1,n^2)$ and $(n^1,n^1,n^1)$.  They come with
the coefficients of $1/3,-1/4,-1/4$ and $1/6$ respectively.  These
coefficients sum up to zero, removing this unphysical state from the
partition function.  Only when $n_1,..,n_m$ are pairwise different and
when  $r_1, ..., r_m$  are equal to one we have a physicial
state.

The  coefficients in the expansion are purely combinatorial
and  have been worked out in \cite{Kostov:2018ckg}
\begin{align}
Z_{ab}(R,L)=\sum_{m=0}^\infty\frac{(-1)^m}{m!}\;\sum_{0\leq
n_1,...,n_m}\prod_{j=1}^m(1-\delta_{n_j,0})\sum_{1\leq
r_1,...,r_m}\frac{(-1)^{r_1+....+r_m}}{r_1....r_m}e^{-RE(n_1^{r_1},...,n_N^{r_m})}.\label{sum-multiplicity}
\end{align}
The  rapidities $u_1, \dots, u_m$ 
of  a generalised Bethe states $(n_1^{r_1},...,n_m^{r_m})$ satisfy the  
Bethe equations
\be
\phi_j=2\pi n_j, \quad j=1,  \dots, m\, ,
\ee
where the scattering phases $\phi _j= \phi _j(u_1,\dots, u_m)$ are 
  defined  by  
%
\begin{align}
e^{ i \phi_j}\equiv 
 e^{2i p(u_j)L } \times R_a(u_j)R_b(u_j)\times
(e^{i\pi}S(u_j,-u_j))^{r_j-1}\times\prod_{k\neq
j}^m(S(u_j,u_k)S(u_j,-u_k))^{r_k}.
\label{Bethe-equations-multiplicity}
\end{align}

\subsection{From mode numbers to rapidities}

In the large $L$ limit, we can replace a discrete sum over mode
numbers $n$ by a continuum integral over variables $\phi$
\begin{align*}
\sum_{0\leq n_1,,,,n_m}=\int_0^\infty\frac{d\phi_1}{2\pi}...\int_0^\infty
\frac{d\phi_m}{2\pi}+\mathcal{O}(e^{-L}).
\end{align*}
We can then use equation \eqref{Bethe-equations-multiplicity} to pass
from $(\phi_1,...,\phi_m)$ to rapidity variables $(u_1,...,u_m)$.  The
only subtle point compared with the periodic case is the factor
excluding the mode numbers $n_j=0$ from the sum
\eqref{sum-multiplicity}
\begin{align}
\sum_{0\leq
n_1,...,n_m}\prod_{j=1}^m(1-\delta_{n_j,0})
=\int_0^\infty\frac{d\phi_1}{2\pi}...\int_0^\infty\frac{d\phi_m}{2\pi}
\prod_{j=1}^m(1-2\pi\delta(\phi_j))+\mathcal{O}(e^{-L}).
\end{align}
We would like to incorporate this factor into the Jacobian matrix
$\partial_u\phi$.  We can do this by first expanding the product as a
sum over subsets $\alpha\subset\lbrace 1,2,...,m\rbrace$, 
\begin{align*}
\int_0^\infty\frac{d\phi_1}{2\pi}...\int_0^\infty\frac{d\phi_m}{2\pi}\sum_{\alpha}(-2\pi)^{|\alpha|}\delta(\phi_\alpha)=&\sum_{\alpha}\prod_{j=1}^m\int_0^\infty\frac{du_j}{2\pi}\bigg[\frac{\partial\phi}{\partial u}\bigg]_{  \alpha,  \alpha}(-2\pi)^{|\alpha|}\delta(u_\alpha)\\
=&\prod_{j=1}^m\int_0^\infty\frac{du_j}{2\pi}\det\bigg[\frac{\partial\phi}{\partial
u}-2\pi\delta(u)\bigg].
\end{align*}
Here   $[\partial\phi/\partial u]_{  \alpha,  \alpha}$ denotes the diagonal minor of the Jacobian matrix obtained by deleting its $\alpha$-rows and $\alpha$-columns. The sum over subsets is the the expansion of the   determinant of a  sum of two matrices.
Hence  the unphysical state at $u=0$ can be eliminated by
adding a term $-2\pi\delta(u)$ to the diagonal elements of the
Jacobian matrix when we change variables from $\phi$ to $u$,
\begin{align}
&G_{jk}(u_1^{r_1},...,u_m^{r_m})\equiv\partial_{u_k}\phi_j-2\pi\delta(u_j)\delta_{jk}\nonumber\\
&=\big[D_{ab}(u_j)+2r_jK(u_j,-u_j)+\sum_{l\neq
j}^mr_l(K(u_j,u_l)+K(u_j,-u_l))\big]\delta_{jk}
\nonumber\\
&-r_k[K(u_k,u_j)-K(u_k,-u_j)] \, (1- \d_{jk}), \quad \forall j,k=1,2,...,m,
\label{Gaudin-multiplicity}
\end{align}
where
\begin{align}
D_{ab}(u)\equiv 2Lp'(u)+\Theta_a(u)+\Theta_b(u).\label{sigma-term}
\end{align}
with $\Theta_a$, $\Theta_b$ defined in \eqref{Omega-term}. In order to apply the matrix-tree theorem, we consider the following matrix
\begin{align}
\hat{G}_{jk}\equiv
r_kG_{kj}=&\big[r_jD_{ab}(u_j,r_j)+2r_j^2\bar{K}_{jj}+\sum_{l\neq
j}^mr_jr_l(K_{jl}+\bar{K}_{jl})\big]\delta_{jk}\nonumber\\
&-r_jr_k(K_{jk}-\bar{K}_{jk})\, (1- \d_{jk}),\quad \forall j,k=1,2,...,m,
\label{Gaudin-hat-multiplicity}
\end{align}
where we have used the notation 
\be
K_{jk}=K(u_j,u_k) , \quad 
 \bar{K}_{jk}=K(u_j,-u_k)=K(-u_j,u_k) . 
 \ee
  In terms of this matrix, the
partition function is written as
\begin{align}
Z_{ab}(R,L)=\sum_{m=0}^\infty\frac{(-1)^m}{m!}\sum_{1\leq
r_1,...,r_m}\prod_{j=1}^m\int_0^\infty\frac{du_j}{2\pi}\frac{(-1)^{r_j}}{r_j^2}
e^{-r_jRE(u_j)}\det
\hat{G}(u_1^{r_1},...,u_m^{r_m}).\label{partition-Gaudin-hat}
\end{align}

\section{Partition function as a sum over graphs}
\label{sec:4}

\subsection{Matrix-tree theorem}
\label{matrix-tree}

The matrix-tree theorem for signed graphs
\cite{Chaiken82acombinatorial} allows us to write the determinant of
the matrix \eqref{Gaudin-hat-multiplicity} as a sum over graphs.  This
theorem as stated in \cite{Chaiken82acombinatorial} is quite technical
and we provide a brief formulation in the following together with two proofs, one combinatorial and one field-theoretical in the appendices.

 First, let us define
\begin{equation}
\begin{aligned}
K^\pm_{jk}&= K_{jk}\pm\bar{K}_{jk}.\label{K-def}
\end{aligned}
\end{equation}
Then the Gaudin-like  matrix \eqref{Gaudin-hat-multiplicity} takes the form
($j,k=1,2,...,m$)
\begin{align}
\hat{G}_{jk}&=\big[r_jD_{ab}(u_j)+r_j^2(K^+_{jj}-K^-_{jj})+\sum_{l\neq
j}^mr_jr_l\, K^+_{jl}\big]\delta_{jk}-r_jr_k\, K^-_{jk} (1-\d_{jk}).
\label{Laplacian-new}
\end{align}
The determinant of this matrix can be written as a sum over all
(not necessarily connected) graphs $\CF$ having exactly  $m$ vertices
labeled by  $v_j$ with $ j=1,..., m$ and two types of edges, positive and negative,  which we denote by   $\ell_{jk}^\pm \equiv  \langle v_j\to v_k\rangle^\pm  $.  
The connected component of
each graph is either:

 \begin{figure}
	 \begin{minipage}[t][][b]{0.29\linewidth}
            \centering
            \includegraphics[width=3.9 cm]{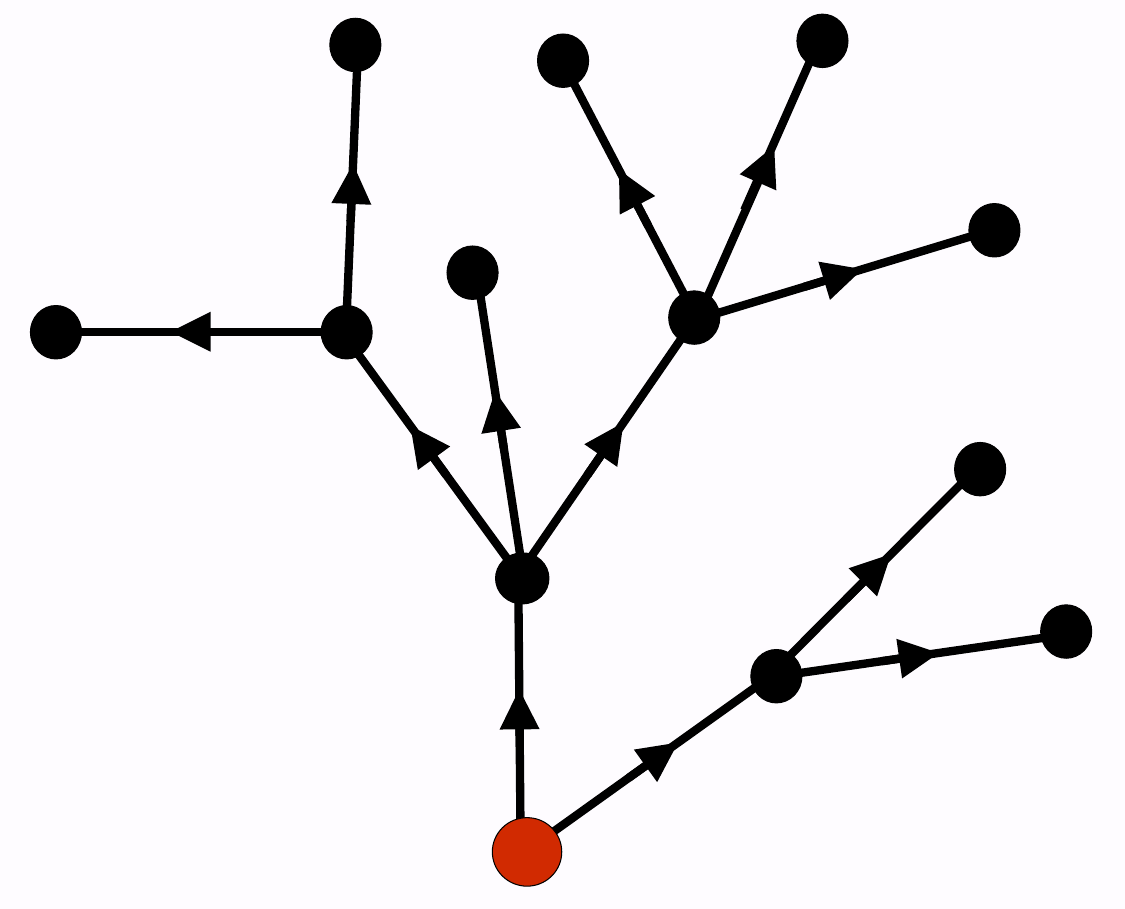}
\caption{ \small    A tree with $K^+$ edges.   }
\label{tree-K+}
         \end{minipage}%
         \hspace{0.8cm}%
                \begin{minipage}[t]{0.65\linewidth}
           \centering
\begin{subfigure}{0.40\textwidth}
\centering
\includegraphics[width=0.9\linewidth]{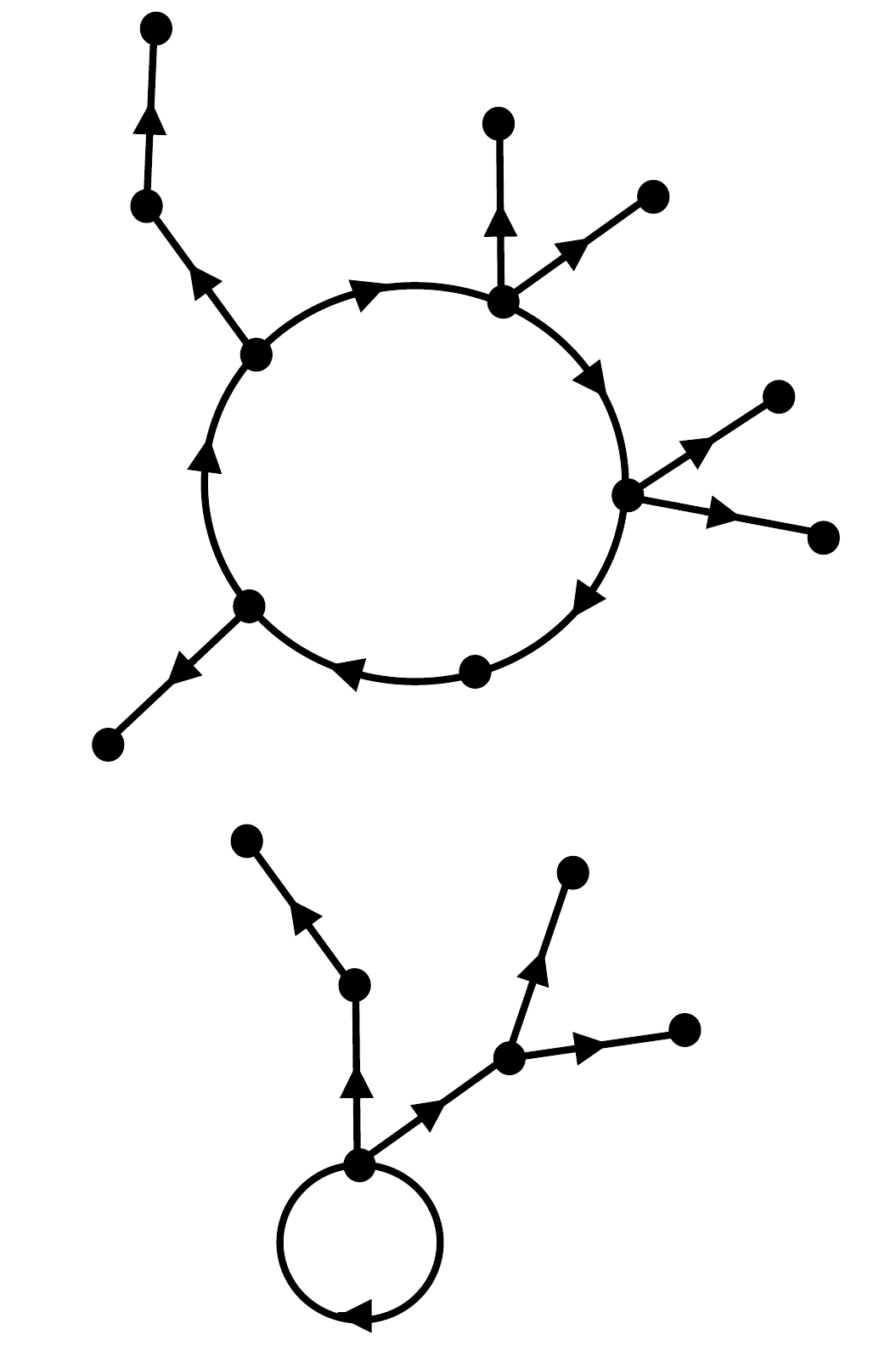}
\caption{$K^+$ loops}
\label{loop-K+}
\end{subfigure}
\hskip 5mm 
 \begin{subfigure}{0.49\textwidth}
 \centering
\includegraphics[width=0.7\linewidth]{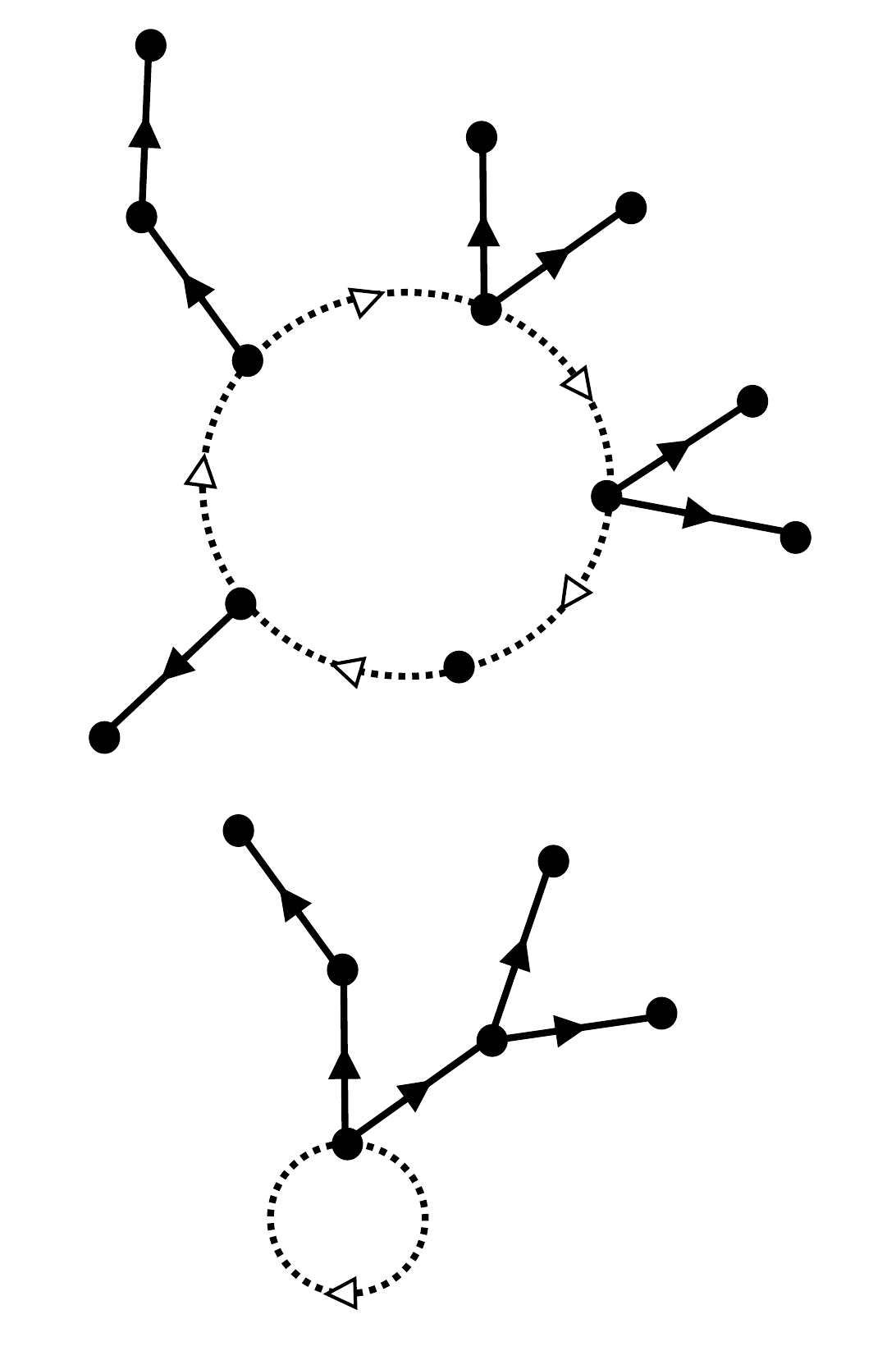}
\caption{$K^-$ loops}
\label{loop-K-}
\end{subfigure}
                 \caption{ \small   Examples of loops with out-growing trees. $K^-$ propagators are drawn as dashed lines. They appear only in a loop and such loop comes with a factor of $-1$. Tadpoles are loops of length $1$.}  
               \label{loops}   \end{minipage}
     \end{figure}

\begin{itemize}
\item A rooted directed tree with  
 only positive edges $\ell_{kl}^+= \langle v_k\to v_l\rangle^+  $
 oriented  so that the edge points to the vertex which is farther from  the root, as 
shown in fig. \ref{tree-K+}.
 The weight of such a tree is a product of 
a factor  $r_jD_{ab}(u_j)$ associated with the root  $v_j$ and  factors
$r_lr_kK^+_{lk}$ associated with its  edges $\ell_{kl}^+$.


\item A positive   (fig.  \ref{loop-K+}) or a  negative  (fig.  \ref{loop-K-}) oriented  cycle 
with outgrowing trees.
A  positive/negatice  loop   is an oriented  cycle (including tadpoles which are cycles of length 1) entirely made of
 positive/negative  edges having the same orientation.  The outgrowing trees consist of positive  edges
only.   The weight of a loop with outgrowing trees is
the product of the weights of its edges, with 
the weight of an edge $ \ell_{kl}^\pm $  
given by 
$r_lr_kK^\pm_{lk}$.  In addition, a negative  loop carries an extra minus sign.  This is why we will call  the   positive   loops
  bosonic  and the negative  loops fermionic.
  \end{itemize}
%

%
Summarising, we write the determinant of  the matrix \re{Laplacian-new} 
as
\be
\begin{split}
\det \hat G_{jk}&=
\sum_\CF W[\CF],
\\
W[\CF]&= (-1)^{\#\text{negative loops}}
\prod_{v_j\in \text{roots}}r_jD_{ab}(u_j)\
\prod  _{\ell_{kl}^\pm \in\text{edges}}   
r_lr_kK^\pm(u_l, u_k),
\la{graphexpansion}
\end{split}
\ee
with $K^\pm(u,v) = K(u,v) \pm K(u,-v)$.
Equation \re{graphexpansion} allows us to express the Jacobian for 
the integration measure as a sum over graphs whose 
weights depend only on the ``coordinates''  $\{ u_j, r_j\}$ of
its  vertices.
For a periodic system $K^+=K^-$ and the two families of loops cancel
each other, leaving only trees in the expansion of the Gaudin matrix
\cite{Kostov:2018ckg}.

\subsection{Graph  expansion of the partition function}

Applying the matrix-tree theorem for each term in the series
\eqref{partition-Gaudin-hat}, we obtain a graph expansion for the
partition function
\begin{align}
Z_{ab}(R,L)=\sum_{m=0}^\infty\frac{(-1)^m}{m!}\sum_{1\leq r_1,...,r_m}\prod_{j=1}^m\int_0^\infty\frac{du_j}{2\pi}\frac{(-1)^{r_j}}{r_j^2} e^{-r_jRE(u_j)}\sum_{\CF}W[\CF],
\end{align}
where the last sum runs over all graphs $\CF$  with  $m$ vertices as
constructed above.

The next step is to invert the order of the sum over graphs and the
integral/sum over the coordinates $\{u_j,r_j\}$ assigned to the
vertices.  As a result we obtain a sum over the ensemble of abstract
oriented tree/loop graphs, with their symmetry factors, embedded in
the space $\mathbb{R}^+\times\mathbb{N}$ where the coordinates $u,r$
of the vertices take values.  The embedding is free, in the sense that
the sum over the positions of the vertices is taken without
restriction.  As a result, the sum over the embedded  graphs is
the exponential of the sum over connected ones.  One can think of
these graphs as Feynman diagrams obtained by applying the Feynman
rules in Fig. \ref{Feynmp}.

The Feynman rules comprise  there kinds of vertices: "root" vertices with only
outgoing bosons, "bosonic" vertices with one incoming boson and an
arbitrary number of outgoing bosons, "fermionic" vertices
with one incoming and one outgoing fermion, together with an
arbitrary number of outgoing bosons. The connected diagrams built from
these vertices are either trees (figure \ref{tree-K+}) or bosonic loops
 (fig.  \ref{loop-K+}) or fermionic loops  (fig.  \ref{loop-K-}).

The free energy is a sum over these graphs, 
\begin{align}
\log Z_{ab}(R,L)=\int_0^\infty\frac{du}{2\pi}D_{ab}(u)\sum_{r\geq
1}rY_r(u)+\sum_{n\geq 1}\mathcal{C}^\pm_n.\label{free-energy}
\end{align}
In this expression, $Y_r(u)$ denotes the  sum of  over all 
trees   rooted at the point $(u,r)$ and $\mathcal{C}^\pm_n$ is the
  sum  over the  Feynman  graphs  having
  a  bosonic/fermionic  loop  of length $n$.  We have defined
$Y_r(u)$ in such a way that the  all vertices with $r$ 
outgoing lines, including the root, have the same weight.

 %

\begin{figure} 
 \be\no
 \boxed{\begin{split} \la{Feynmp} 
  \imineq{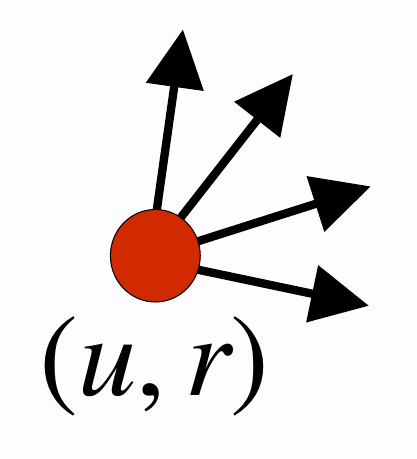}{9} \  \qquad  &=
   \  \  \  rD_{ab}(u)
    {(-1)^{r-1} \over r^2} \ e^{-r R E(u)}
\\
 \imineq{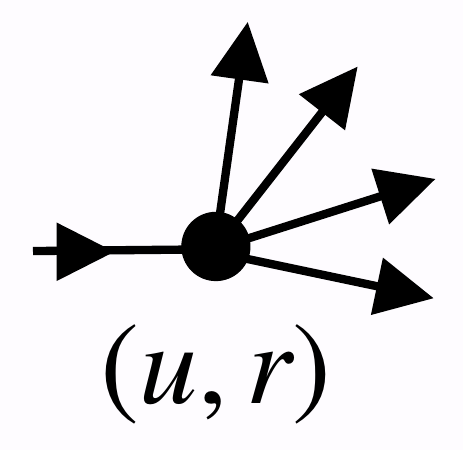}{9}\quad=\quad\imineq{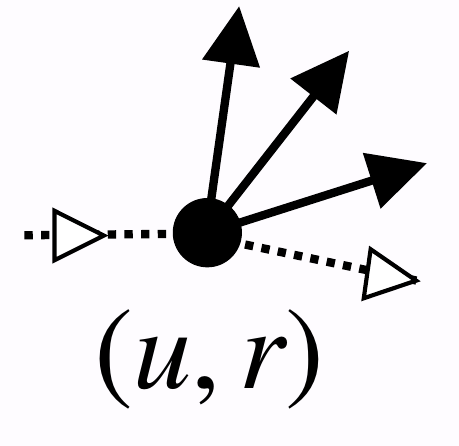}{9}  \ \qquad  &= \  \  \   {(-1)^{r-1} \over r^2} \ e^{-r R E(u)}
\\
\imineq{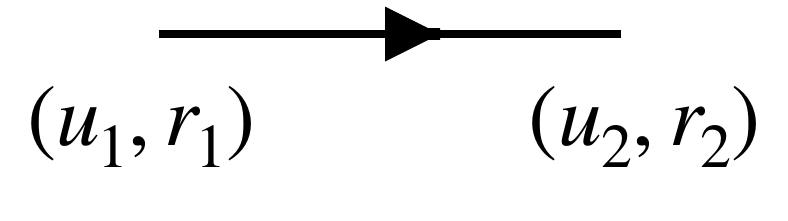}{6} \ \quad   &= \ \ \ r_1 r_2 K^+(u_2, u_1)
\\
\imineq{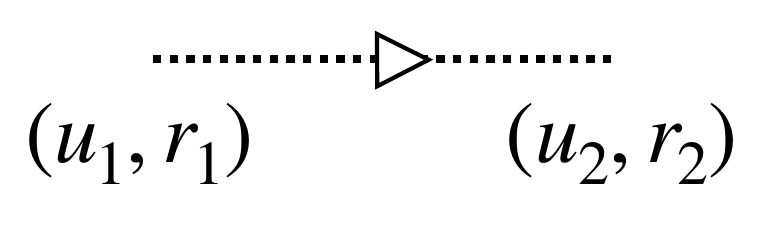}{6}\ \quad   &= \ \ \ r_1 r_2 K^-(u_2, u_1)
\\
  \imineq{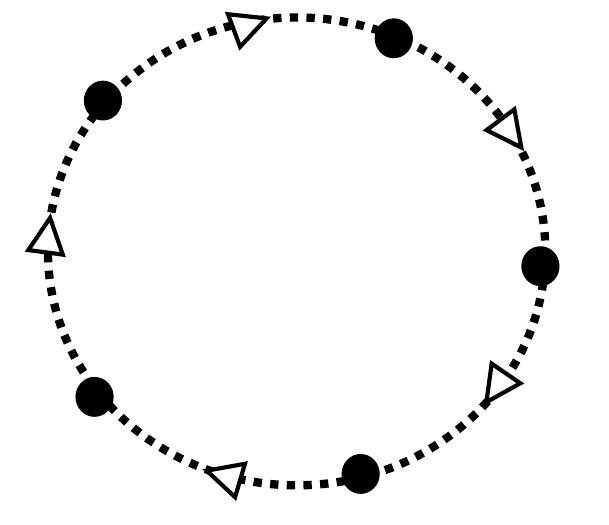}{11} \quad &= \ \ \ -1
 \end{split} 
 }\no
 \ee
  \caption{ \small  The Feynman rules for the partition function. The vertex 
  labeled by $(u,r)$ has $r$ outgoing lines.}
  \end{figure}

\section{Summing up the connected graphs: the exact $g$-function}
\label{sec:5}

\subsection{The tree contribution}

\begin{figure}[h]
\centering
\includegraphics[width=12cm]{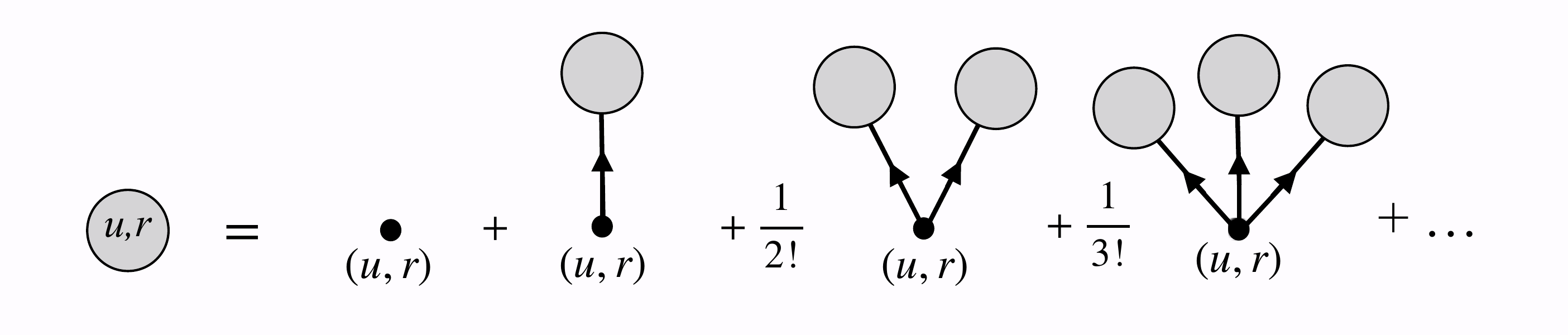}
\caption{\small The combinatorial structure of trees with a fixed root.}
\label{iterative-equation}
\end{figure}

In this section, we analyze the part of free energy
\eqref{free-energy} that comes from the tree-diagrams
\begin{gather}
\log Z_{ab}(R,L)^\textnormal{trees}=\int_0^{+\infty}\frac{du}{2\pi}D_{ab}(u)\sum_{r\geq 1}rY_r(u),
\\
\text{where}\quad Y_r(u)=\sum\vcenter{\hbox{\includegraphics[width=2.5cm]{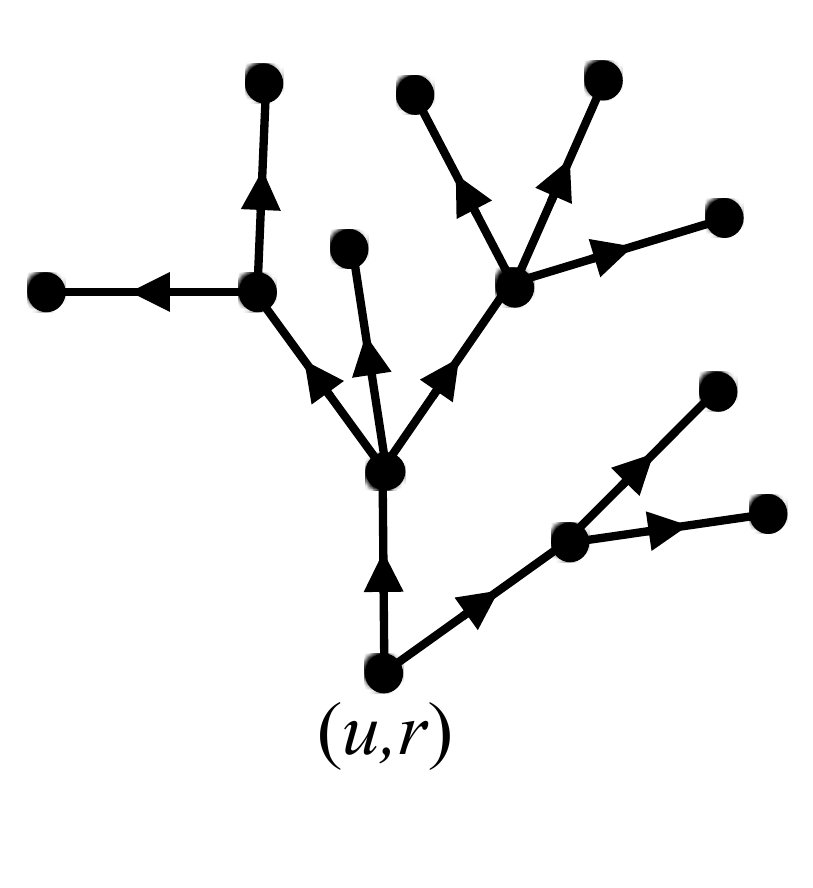}}}\;\equiv \vcenter{\hbox{\includegraphics[width=0.70cm]{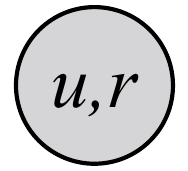}}}
.\label{free-energy-trees}
\end{gather}
Being the generating function for directed trees rooted at $(u,r)$,
$Y_r(u)$ obeys a simple equation
\begin{align}
Y_r(u)&=\frac{(-1)^{r-1}}{r^2}e^{-rRE(u)}\sum_{n=0}^\infty\frac{1}{n!}
\bigg(\sum_{s=1}^\infty\int_0^{+\infty}\frac{dv}{2\pi}
srK^+(v,u)Y_s(v)\bigg)^n\nonumber\\
&=\frac{(-1)^{r-1}}{r^2}\bigg[ e^{-RE(u)}\exp\sum_{s=0}^\infty\int_0^{+\infty}
\frac{dv}{2\pi}  K^+(v,u)sY_s(v)\bigg] ^r,\label{TBA-1}
\end{align}
This equation can be understood diagramatically as in figure
\ref{iterative-equation}.

\noindent
In particular, we have for $r=1$
\begin{align}
Y_1(u)= e^{-RE(u)}\exp\sum_{s=0}^\infty\int_0^{+\infty}\frac{dv}{2\pi}
K^+(v,u)sY_s(v).\label{Y1}
\end{align}
By replacing \eqref{Y1} into \eqref{TBA-1}, we can express $Y_r$ in
terms of $Y_1$ for arbitrary $r\geq 1$
\begin{align}
Y_r(u)=\frac{(-1)^{r-1}}{r^2}Y_1(u)^r.\label{relation}
\end{align}
This allows us to rewrite \eqref{Y1} as a closed equation for $Y_1$
\begin{gather*}
 Y_1(u)=e^{-RE(u)}\exp\int_0^{+\infty}\frac{dv}{2\pi} 
K^+(v,u)\log[1+Y_1(v)].\label{TBA-2}
\end{gather*}
This integral can be extended to the real axis by using the parity of
the kernel $K^+(v,u)=K(v,u)+K(-v,u)$ and by defining $Y_1(-u)=Y_1(u)$
\begin{gather*}
Y_1(u)=e^{-RE(u)}\exp\int_{-\infty}^{+\infty}\frac{dv}{2\pi}  K(v,u)\log[1+Y_1(v)],\label{TBA-periodic}
\end{gather*}
This is nothing but the TBA equation for a periodic system at inverse temperature $R$. In particular, the periodic partition function can be written in terms of $Y_1$
\begin{align}
\log Z(R,L)=L\int_{-\infty}^{+\infty}\frac{du}{2\pi}p'(u)\log[1+Y_1(u)].\label{free-energy-periodic}
\end{align}
Similarly, we can also extend the domain of integration in
\eqref{free-energy-trees} to the real axis, using the parity of
$D_{ab}(u,r)$ and $Y_1$.  By subtracting the periodic free energy
\eqref{free-energy-periodic} from the tree part of the free energy
\eqref{free-energy-trees}, we obtain the tree contribution to
$g$-function
\begin{align}
\mathcal{F}_{ab}(R)^\textnormal{trees}=\frac{1}{2}\int_{-\infty}^{+\infty}\frac{du}{2\pi}[\Theta_a(u)+\Theta_b(u)]\log[1+Y_1(u)].\label{free-energy-trees-3}
\end{align}

\subsection{Loop contribution}

Now we turn to the sum over loops and show that they fill the missing
part \cite{Dorey:2004xk} of the $g$-function \eqref{$g$-function}.  Let us define
\begin{align}
\mathcal{F}_{ab}(R,L)^\textnormal{loops}=\sum_{n\geq
1}\mathcal{C}_n^\pm\label{free-energy-cycles}.
\end{align}
For each $n\geq 1$, $\mathcal{C}_n^\pm$ is the partition sum of
$K^\pm$ loops of length $n$ with the trees growing out of these loops which can
be summed separately
\begin{align*}
\mathcal{C}_n^\pm=\frac{\pm 1}{n}\sum_{1\leq
r_1,...,r_n}\int\limits_0^\infty\frac{du_1}{2\pi}...\int\limits_0^\infty
\frac{du_n}{2\pi}Y_{r_1}(u_1)....Y_{r_n}(u_n)r_2r_1K^\pm(u_2,u_1)....r_1r_nK^\pm(u_1,u_n).
\end{align*}
In this expression, the sign comes from fermion loop and $1/n$ is the
usual loop symmetry factor.

\begin{figure}[h]
\centering
\includegraphics[width=5.0cm]{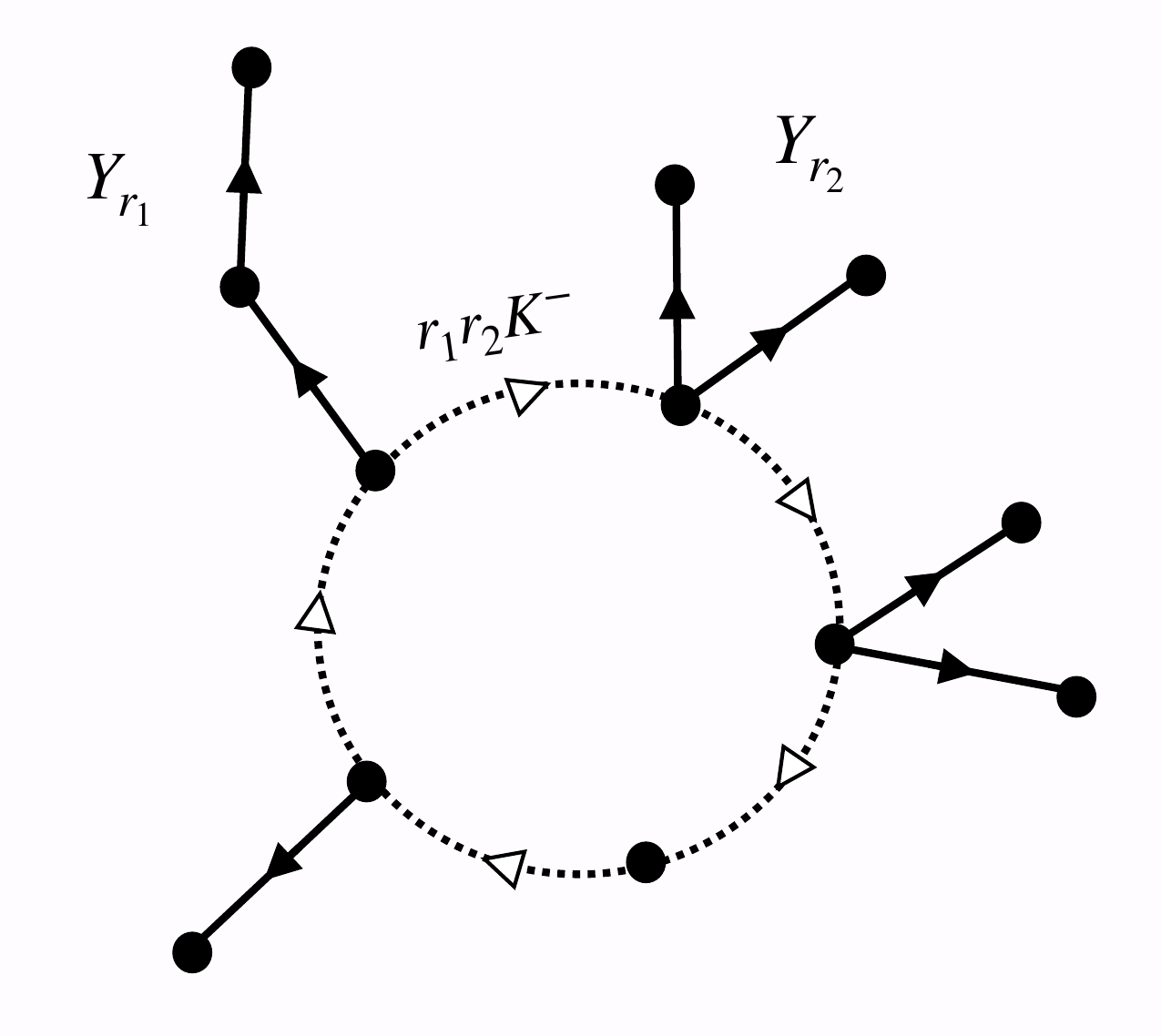}
\caption{\small A $K^-$ loop connecting $n$ points $(u_j,r_j)$. Trees growing out of a vertex $(u_j,r_j)$ sum up to the corresponding $Y_r(u)$ function. The factor $r_{j-1}r_j$ of the propagator from $(u_{j-1},r_{j-1})$ to $(u_j,r_j)$ can be pulled into the adjacent trees.
Taking the sum over all $r_j$  results in the Fermi-Dirac factor $Y_1(u_j)/(1+Y_1(u_j))$ at each vertex $j$.}
\end{figure}

We can use the relation \eqref{relation} to carry out the sum over
$r$
\begin{align*}
\sum_{r\geq 1}r^2Y_r(u)=\frac{Y_1(u)}{1+Y_1(u)}.
\end{align*} 
It follows that
\begin{align}
\mathcal{C}_n^\pm =\frac{\pm 1}{n}\tr (\hat{K}^\pm)^n.
\end{align}
where the kernels $\hat{K}^\pm$ are defined in \eqref{Q}.  The loop
contribution is therefore given by
\begin{align}
\mathcal{F}_{ab}(R)^{\textnormal{loops}}=\log\det\frac{1-\hat{K}^-}{1-\hat{K}^+}.\label{free-energy-cycles-2}
\end{align}

 The $g$-function is obtained by combining \eqref{free-energy-trees-3}
 and \eqref{free-energy-cycles-2} and set $a=b$
\begin{align}
\log g_a(R)=&\frac{1}{2}\big[\mathcal{F}_{aa}^\textnormal{trees}(R)+\mathcal{F}_{aa}^\textnormal{loops}(R)\big]\nonumber\\
=&\frac{1}{2}\int_{-\infty}^{+\infty}\frac{du}{2\pi}\Theta_{a}(u)\log[1+Y_1(u)]+\frac{1}{2}\log\det\frac{1-\hat{K}^-}{1-\hat{K}^+}.\label{final}
\end{align}
\subsection{Excited state g-function}
In this section, we derive the excited state g-function. This quantity can be regarded as the normalized overlap between the boundary state and an excited bulk eigenstate
\begin{align}
g_a^{\psi}=\frac{\langle B_a|\psi\rangle}{\sqrt{\langle \psi|\psi\rangle}}.\label{overlap-excited}
\end{align}
By setting $|\psi\rangle$ to the ground state $|\psi_0\rangle$, we recover the definition \eqref{overlap-ground} of the g-function. We restrict our computation to the case where $|\psi\rangle$ is of the form $|\pm w_1,\pm w_2,...,\pm w_N\rangle$. We also assume for simplicity  that the scattering matrix is a function of the difference of rapidities (relativistic invariance).

First, let us briefly summarize the excited state TBA equations for a periodic system, following \cite{Kostov:2018ckg}. We consider a torus with one large dimension $L$ (physical volume) and a finite dimension $R$ (mirror volume). A mirror state $|\vartheta\rangle=|v_1,...,v_N\rangle$ propagates along the $L$ direction. Note that the mirror-physics convention is in reverse order compared to \cite{Kostov:2018ckg}. The Boltzmann weight of a physical particle is dressed by the interaction with these mirror particles
\begin{align}
Y_\vartheta^\circ(u)=e^{-RE(u)}\prod_{j=1}^N S(u-v_j+i\pi/2),
\end{align}
To compute the energy of the mirror state $|\vartheta\rangle$, we have to add to the free energy the contribution from the mirror particles that go directly to the opposite edge without scattering
\begin{align}
E(\vartheta)=-\frac{1}{L}\log Z(R,L,\vartheta)=\sum_{j=1}^N {E}(v_j)-\int_{-\infty}^{+\infty}\frac{du}{2\pi}p'(u)\log(1+Y_\vartheta(u)),\label{excited-energy}
\end{align}
where $Y_\vartheta$ solves for the excited state TBA equation
\begin{align}
Y_\vartheta(u)=Y_\vartheta^\circ(u)\exp\big[\int_{-\infty}^{+\infty} \frac{dw}{2\pi} K(w,u)\log (1+Y_\vartheta(w))\big].\label{excited-TBA}
\end{align}
The on-shell condition for the state $|\vartheta\rangle$ is  obtained by transforming a mirror particle of rapidity $v_j$ to a physical particle of rapidity $v_j-i\pi/2$. The relative factor between the two ways of computing the partition function is $-Y(v_j-i\pi/2)$. This leads to the finite volume Bethe equations
\begin{align}
Y(v_j-i\pi/2)=-1,\quad j=1,2,...,N.\label{excited-Bethe}
\end{align}

Now let us return  to the excited state g-function \eqref{overlap-excited}. We repeat the same exercise for a long cylinder of length $L$ and radius $R$ with two boundaries $a$ and $b$ together with a state $|\psi\rangle=|\pm w_1,\pm w_2,...,\pm w_N\rangle$ propagating in the $L$ direction. We denote the partition function in this case by $Z_{ab}(R,L,\psi)$.

The idea is, if we can identify the excited energy \eqref{excited-energy} with the extensive part of the partition function $Z_{ab}(R,L,\psi)$ when $|\psi\rangle\equiv |\vartheta\rangle$, then the rest (intensive part) gives us the excited g-function corresponding to $|\psi\rangle$
\begin{align}
g_a^\psi g_b^\psi=\frac{Z_{ab}(R,L,\psi)}{Z(R,L,\psi)}.
\end{align}

To compute $Z_{ab}(R,L,\psi)$ we perform the sum over eigenstates of the physical Hamiltonian with boundary $H_{ab}$. The procedure is similar to that of ground-state g-function: we obtain a sum over trees and loops. The only difference is the Feynman rule for the vertices
\begin{align}
e^{-RE(u)}\to e^{-RE(u)}\prod_{j=1}^NS(u-w_j+i\pi/2)S(u+w_j+i\pi/2)\equiv\tilde{Y}_\psi^\circ(u)
\end{align}

In particular, the extensive part of the partition function $Z_{ab}(R,L,\psi)$ is given by
\begin{align}
\log Z_{ab}(R,L,\psi)^\textnormal{extensive}=-L\sum_{j=1}^N {E}(\pm w_j)+2L\int_{0}^\infty\frac{du}{2\pi}p'(u)\log(1+\tilde{Y}_\psi(u))\label{excited-energy-boundary}
\end{align}
where $\tilde{Y}_\psi(u)$ being the sum of trees rooted at vertex $u$ now satisfies the equation 
\begin{align}
\tilde{Y}_\psi(u)=\tilde{Y}_\psi^\circ(u)\exp\big[\int_{0}^{+\infty} \frac{dw}{2\pi} K^+(w,u)\log (1+\tilde{Y}_\psi(w))\big].\label{excited-TBA-boundary}
\end{align}
As a consequence of the  crossing symmetry $S(u) = S( i\pi-u)$
we have the identity 
\begin{align}
S(u-w_j+i\pi/2)S(u+w_j+i\pi/2)=S(-u-w_j+i\pi/2)S(-u+w_j+i\pi/2)\label{parity-excited}
\end{align}
 which means that  the function  $\tilde{Y}_\psi^\circ(u)$ is an even function of $u$.
Therefore we can extend $\tilde{Y}_\psi$ to the real axis and identify $\tilde{Y}_\psi$ with $Y_\vartheta$ when $|\psi\rangle =|\vartheta\rangle$. Again we have $\tilde{Y}_\psi(w_j)=\tilde{Y}_\psi(-w_j)=-1$. We conclude that
\begin{align}
\log(g_a^\psi g_b^\psi)(R)=\int_0^\infty\frac{du}{2\pi}[\Theta_a(u)+\Theta_b(u)]\log(1+\tilde{Y}_\psi(u))+\log\det\frac{1-\hat{K}^-_\psi}{1+\hat{K}^+_\psi},\label{excited-g}
\end{align}
where the Fermi-Dirac factor in the kernel $\hat{K}^\pm_\psi$ is now given by $\tilde{Y}_\psi/(1+\tilde{Y}_\psi)$, c.f. eq. \re{Q}.

Our method produces correctly equations \eqref{excited-TBA}, \eqref{excited-Bethe}, \eqref{excited-TBA-boundary} that determine the excited states in both periodic and open case. However in the periodic case our intuitive picture misses a contribution $\sum E(v_j)$ to the excited state energy. Similarly, we believe that the structure of the excited state g-function given in \eqref{excited-g} is correct up to a simple additional contribution. By comparison with the asymptotic g-function derived in \cite{Kormos:2010ae}, this term appears to be the mirror-continued reflection factor. We leave this problem to future investigations.

\section{More than one type of particle}
In this section we generalize our method to theories with multiple species of particle interacting via diagonal bulk scattering and diagonal reflection matrices. This generalization also provides insight on g-function of non-diagonal theories solved by Nested Bethe Ansatz. We present a regularization scheme for these theories and we consider an explicit example in another work \cite{Vu2019}.

For simplicity we consider a theory with two species.  The bulk
scattering matrices and the reflection matrices are denoted by
$S_{pq}$ and $R_{pa},R_{pb}$ for $p,q\in\lbrace 1,2\rbrace$. They are
assumed to satisfy the following properties
\begin{gather}
S_{pp}(u,u)=-1,\nonumber\\
S_{pq}(u,v)S_{qp}(v,u)=R_{pa}(u)R_{pa}(-u)=R_{pb}(u)R_{pb}(-u)=1,\\
S_{pq}(u,-v)S_{qp}(-u,v)=1.\nonumber
\end{gather}
The last property is only needed for system with boundaries.
\subsection{Periodic systems}
An $(N_1+N_2)$-particle state is characterized by a set of rapidities
$|u_{11},...,u_{1N_1},u_{21},...,u_{2N_2}\rangle$.  Particles of the
same type must have different rapidities: $u_{1j}\neq u_{1k}$,
$u_{2j}\neq u_{2k}$.  The Bethe equations for such state read
\begin{equation}
\begin{aligned}
p_1(u_{1j})L+\sum_{k\neq j}^{N_1}-i\log
S_{11}(u_{1j},u_{1k})+\sum_{k=1}^{N_2}-i\log
S_{12}(u_{1j},u_{2k})=\phi_{1j}=2\pi n_{1j},\\
p_2(u_{2j})L+\sum_{k=1}^{N_1}-i\log S_{21}(u_{2j},u_{1k})+\sum_{k\neq j}^{N_2}-i\log S_{22}(u_{2j},u_{2k})=\phi_{2j}=2\pi n_{2j}.
\end{aligned}
\end{equation}
The partition function can be written as a sum runs over two sets of
mode numbers $\textbf{n}_1=n_{11},...,n_{1m_1}$ and
$\textbf{n}_2=n_{21},...,n_{2m_2}$ along with two sets of
multiplicities (wrapping numbers) $\textbf{r}_1=r_{11},...,r_{1m_1}$
and $\textbf{r}_2=r_{21},...,r_{2m_2}$
\begin{align}
Z_{ab}(R,L)=\sum_{\substack{m_1=0\\m_2=0}}^\infty\frac{(-1)^{m_1+m_2}}{m_1!m_2!}\;\sum_{\substack{0\leq \textbf{n}_1,\textbf{n}_2\\1\leq \textbf{r}_1,\textbf{r}_2}}\;\prod_{p=1}^2\prod_{j=1}^{m_p}\frac{(-1)^{r_{pj}}}{r_{pj}}e^{-RE((\textbf{n}_1,\textbf{r}_1),(\textbf{n}_2,\textbf{r}_2))}.\label{sum-multiplicity-su(3)-periodic}
\end{align}
The mode numbers
$((\textbf{n}_1,\textbf{r}_1),(\textbf{n}_2,\textbf{r}_2))$ are
related to the rapidities
$((\textbf{u}_1,\textbf{r}_1),(\textbf{u}_2,\textbf{r}_2))$ through
Bethe equations with multiplicities
\begin{align*}
p_1(u_{1j})L+\sum_{k\neq j}^{m_1}-ir_{1k}\log
S_{11}(u_{1j},u_{1k})+\sum_{k=1}^{m_2}-ir_{2k}\log
S_{12}(u_{1j},u_{2k})=\phi_{1j}=2\pi n_{1j},\\
p_2(u_{2j})L+\sum_{k=1}^{m_1}-ir_{1k}\log S_{21}(u_{2j},u_{1k})+\sum_{k\neq j}^{m_2}-ir_{2k}\log S_{22}(u_{2j},u_{2k})=\phi_{2j}=2\pi n_{2j}.
\end{align*}
The Gaudin matrix has a $2\times 2$ block structure
\begin{gather}
\hat{G}=\begin{pmatrix}
\textbf{r}_1\partial_{u_1}\phi_1&\textbf{r}_1\partial_{u_2}\phi_1\\
\textbf{r}_2\partial_{u_1}\phi_2&\textbf{r}_2\partial_{u_2}\phi_2
\end{pmatrix}^t=\begin{pmatrix}
\hat{A}&\hat{B}\\
\hat{C}&\hat{D}
\end{pmatrix}\label{Gaudin-su(3)-periodic}
\end{gather}
The explicit expressions of each block are
\begin{gather*}
\hat{A}_{jk}=\delta_{jk}[r_{1j}Lp'_{1j}+\sum_{l\neq
j}^{m_1}r_{1j}r_{1l}K_{11}^{jl}+\sum_{l=1}^{m_2}
r_{1j}r_{2l}K_{12}^{jl}]-r_{1j}r_{1k}K_{11}^{jk},\\
\hat{B}_{jk}=-r_{1j}r_{2k}K_{12}^{jk},\quad \hat{C}_{jk}=-r_{2j}r_{1k}K_{21}^{jk},\\
\hat{D}_{jk}=\delta_{jk}[r_{2j}Lp'_{2j}+\sum_{l=1}^{m_1}r_{2j}r_{1l}K_{21}^{jl}+\sum_{l\neq
j}^{m_2}r_{2j}r_{2l}K_{22}^{jl}]-r_{2j}r_{2k}K_{22}^{jk}.
\end{gather*}
The partition function can be written in terms of the determinant of this matrix
\begin{align}
Z(R,L)=\sum_{\substack{m_1=0\\m_2=0}}^\infty\frac{1}{m_1!m_2!}
\sum_{\textbf{r}_1,\textbf{r}_2}\prod_{p=1}^2\prod_{j=1}^{m_p}
\int_{-\infty}^{+\infty}\frac{du_{pj}}{2\pi}\frac{(-1)^{r_{pj}-1}}{r_{pj}^2}e^{-r_{pj}RE_p(u_{pj})
}\det\hat{G}.
\end{align}

We apply the matrix-tree theorem for the matrix $\hat{G}$ and obtain a
tree expansion of the free energy.  Each vertex now carries an index
$p\in\lbrace 1,2\rbrace$ to indicate what type of particle it stands
for.  A branch going from vertex of type $p$ to vertex of type $q$ has a weight of
$r_qr_pK_{qp}$. We represent vertices of type 1 by a disk and those of type 2
by a circle.

Let us denote by $Y_{pr}(u)$ the sum over all the trees rooted at
$(u,r)$ of type $p$.  The free energy is given by
\begin{align}
\log Z(R,L)=L\int_{-\infty}^{+\infty}\frac{du}{2\pi}\bigg[ p_1'(u)\sum_rrY_{1r}(u)+p_2'(u)\sum_rY_{2r}(u)\bigg].
\end{align}
\medskip \medskip 
 \be
 \boxed{\begin{split} \la{su(3)-Feynmp} 
   \imineq{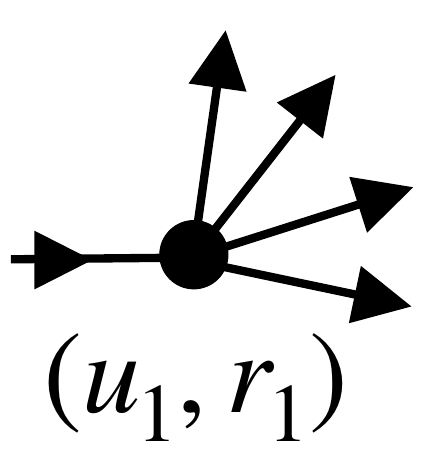}{10} \ \quad  &= \  \  \   {(-1)^{r_1-1} \over r_1^2} \ e^{-r_1 R E_1(u)}
\\
\imineq{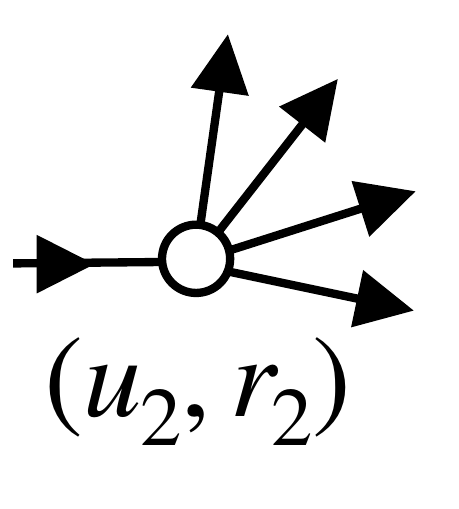}{10.5} \ \quad   &= \ \ \ {(-1)^{r_2-1} \over r_2^2}\ e^{-r_2 R E_2(u)}
 \end{split} 
 }\ee

The generating functions of the two types of trees are intertwined with each other
\begin{align*}
Y_{1r}(u)&=\frac{(-1)^{r-1}}{r^2}e^{-rRE_1(u)}\exp\bigg[r\int\frac{dv}{2\pi}
\sum_{s}sK_{11}(v,u)Y_{1s}(v)+sK_{21}(v,u)Y_{2s}(v)\bigg],\\
Y_{2r}(u)&=\frac{(-1)^{r-1}}{r^2}e^{-rRE_2(u)}\exp\bigg[r\int\frac{dv}{2\pi}
\sum_ssK_{12}(v,u)Y_{1s}(v)+sK_{22}(v,u)Y_{2s}(v)\bigg].
\end{align*}
\begin{figure}[h]
\centering
\includegraphics[width=15cm]{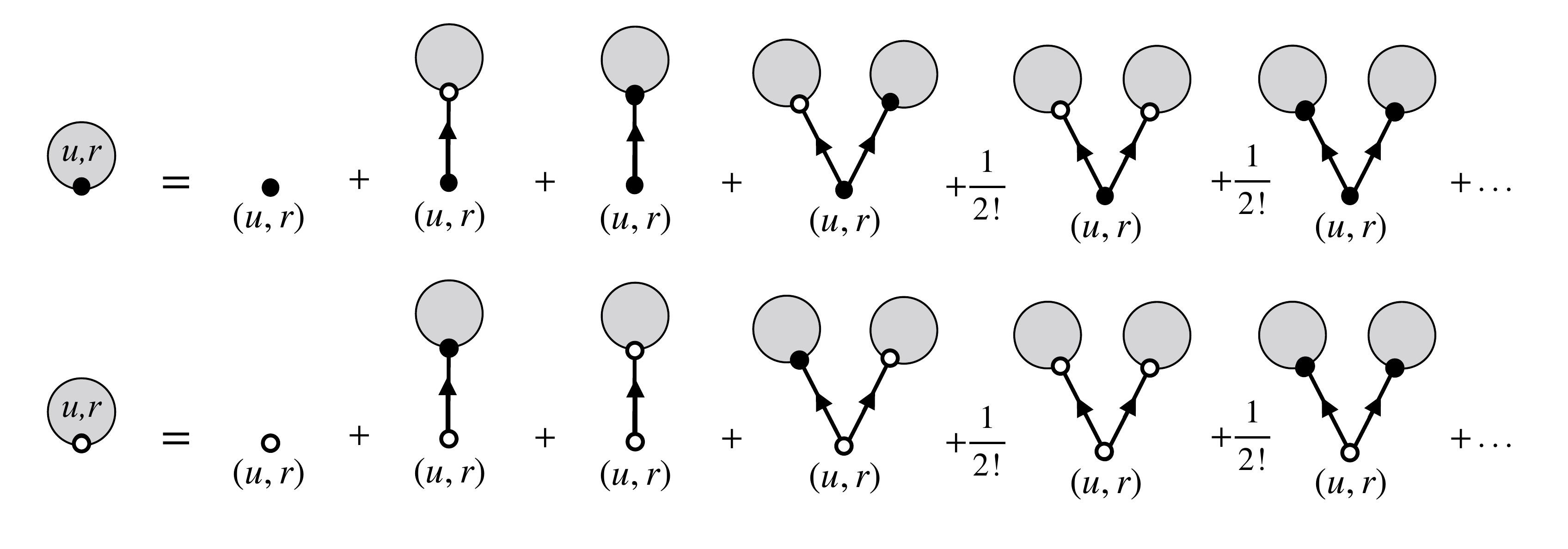}
\caption{The two-species TBA equations}
\end{figure}

In particular, we have
\begin{align}
Y_{1r}(u)=\frac{(-1)^{r-1}}{r^2}Y^r_{11}(u),\quad
Y^r_{2r}(u)=\frac{(-1)^{r-1}}{r^2}Y^r_{21}(u).\label{su(3)-relation}
\end{align}
For simplicity, let us denote $Y_{11}$ and $Y_{21}$ simply by $Y_1$
and $Y_2$.  We recover the TBA equation of two species
\begin{equation}
\begin{aligned}
\log Z(R,L)&=L\int_{-\infty}^{+\infty}\frac{du}{2\pi}\bigg\lbrace p_1'(u)\log[1+Y_1(u)]+p_2'(u)\log[1+Y_2(u)]\bigg\rbrace,\\
Y_1(u)&=e^{-RE_1(u)}\exp\big[K_{11}\star \log(1+Y_1)+K_{21}\star \log(1+Y_2)\big],\\
Y_2(u)&=e^{-RE_2(u)}\exp\big[K_{12}\star\log(1+Y_1)+K_{22}\star\log(1+Y_2)\big].
\end{aligned}\label{su(3)-TBA}
\end{equation}

\subsection{Open systems}

Let us denote  by $R_{1ab}$ and $R_{2ab}$ the reflection factors of the first and second particle.The Bethe equations for the state
$|u_{11},...,u_{1N_1},u_{21},...,u_{2N_2}\rangle$ now read
\begin{equation*}
\begin{aligned}
e^{2ip_1(u_{1j})L}R_{1ab}(u_{1j})\prod_{k\neq
j}^{N_1}S_{11}(u_{1j},u_{1k})S_{11}(u_{1j},-u_{1k})\prod_{k=1}^{N_2}S_{12}(u_{1j},u_{2k})S_{12}(u_{1j},-u_{2k})=1,\\
e^{2ip_2(u_{2j})L}R_{2ab}(u_{2j})\prod_{k=1}^{N_1}
S_{21}(u_{2j},u_{1k})S_{21}(u_{2j},-u_{1k})\prod_{k\neq j}^{N_2}
S_{22}(u_{2j},u_{2k})S_{22}(u_{2j},-u_{2k})=1.
\end{aligned}
\end{equation*}
The rapidities and the mode numbers are taken to be positive.  Similar
to \eqref{sum-multiplicity}, we have
\begin{align}
Z_{ab}(R,L)=\sum_{\substack{m_1=0\\m_2=0}}^\infty\frac{(-1)^{m_1+m_2}}{m_1!m_2!}\;
\sum_{\substack{0\leq
\textbf{n}_1,\textbf{n}_2\\1\leq
\textbf{r}_1,\textbf{r}_2}}\;\prod_{p=1}^2\prod_{j=1}^{m_p}
\frac{(-1)^{r_{pj}}}{r_{pj}}(1-\delta_{n_{pj},0})
e^{-RE((\textbf{n}_1,\textbf{r}_1),(\textbf{n}_2,\textbf{r}_2))}.
\label{sum-multiplicity-su(3)}
\end{align}
The conversion between mode numbers and rapidities under the presence
of multiplicities
\begin{align*}
2p_1(u_{1j})L-i\log\bigg[R_{1ab}(u_{1j})[\bold S_{11}(u_{1j},u_{1j})]^{r_{1j}-1}\prod_{k\neq
j}^{m_1} [\bold S_{11}(u_{1j}, u_{1k})]^{r_{1k}}\prod_{k=1}^{m_2}
[\bold S_{12}(u_{1j},u_{2k})]^{r_{2k}}\bigg]&=\phi_{1j}\\
2p_2(u_{2j})L-i\log\bigg[R_{2ab}(u_{2j})[\bold S_{22}(u_{2j},u_{2j})]^{r_{2j}-1}\prod_{k=1}^{m_1} 
[\bold S_{21}(u_{2j},u_{1k})]^{r_{1k}}\prod_{k\neq j}^{m_2} 
[\bold S_{22}(u_{2j},u_{2k})]^{r_{2k}}\bigg]&=\phi_{2j}
\end{align*}
where we have used the notation
$\bold S_{pq}(u,v)=S_{pq}(u,v)S_{pq}(u,-v)$.  The Gaudin matrix now has
a $2\times 2$ block structure
\begin{align}
&\hat{G}_{ab}=\begin{pmatrix}
r_1[\partial_{u_1}\phi_1-2\pi\delta(u_1)]&r_1\partial_{u_2}\phi_1\\
r_2\partial_{u_1}\phi_2&r_2[\partial_{u_2}\phi_2-2\pi\delta(u_2)]
\end{pmatrix}^t=\begin{pmatrix}
\hat{A}&\hat{B}\\
\hat{C}&\hat{D}.
\end{pmatrix}\label{Gaudin-su(3)-boundary}
\end{align}
The explicit expressions of each block are
\begin{align*}
&\hat{A}_{jk}=\delta_{jk}\bigg[r_{1j}[2Lp_1'(u_{1j})
+\Theta_{1ab}(u_{1j})]+r_{1j}^2(K_{11}^{jj+}-K_{11}^{jj-})+\sum_{l\neq
j}^n r_{1j}r_{1l}K_{11}^{jl+}+\sum_{l=1}^m
r_{1j}r_{2l}K_{12}^{jl+}\bigg]\\
&\qquad\qquad\qquad\qquad\qquad\qquad\qquad\qquad\qquad-r_{1j}r_{1k}K_{11}^{jk-},\\
&\hat{B}_{jk}=-r_{1j}r_{2k}K_{12}^{jk-},\quad \hat{C}_{jk}=-r_{2j}r_{1k}K_{21}^{jk-},\\
&\hat{D}_{jk}=\delta_{jk}\bigg[r_{2j}[2Lp_2'(u_{1j})
+
\Theta_{2ab}(u_{2j})]+r_{2j}^2(K_{22}^{jj+}-K_{22}^{jj-})+\sum_{l=1}^m
r_{2j}r_{1l}K_{21}^{jl+}+\sum_{l\neq j}^m
r_{2j}r_{2l}K_{22}^{jl+}\bigg]\\
&\qquad\qquad\qquad\qquad\qquad\qquad\qquad\qquad\qquad-r_{2j}r_{2k}K_{22}^{jk-}.
\end{align*}
where the notations are
\begin{gather*}
\Theta_{pab}=\Theta_{pa}+\Theta_{pb}, \quad
\Theta_{p\star}(u)=K_{p\star}(u)-K_{pp}(u,-u)-\pi\delta(u),\;\star=a,b\\
K_{pq}^{jk\pm}=K_{pq}(u_{pj},u_{qk})\pm K_{pq}(u_{pj},-u_{qk}),\quad \textnormal{for } p,q\in\lbrace 1,2\rbrace.
\end{gather*}
If we set $\Theta$ to zero and $K^+$ and $K^-$ to equal then we would
recover the Gaudin matrix for the periodic system
\eqref{Gaudin-su(3)-periodic}.  The partition function is written in
terms of the determinant of this matrix
\begin{align}
Z_{ab}(R,L)=\sum_{\substack{m_1=0\\m_2=0}}\frac{1}{m_1!m_2!}\sum_{1\leq\textbf{r}_{1},\textbf{r}_2}\prod_{p=1}^2\prod_{j=1}^{m_p}\int_{-\infty}^{+\infty}\frac{du_{pj}}{2\pi}\frac{(-1)^{r_{pj}-1}}{r_{pj}^2}e^{-r_{pj}RE_p(u_{pj})}\det\hat{G}_{ab}.\label{partition-su(3)-boundary}
\end{align}
The tree contribution to $g$-function is obtained in a similar way as
before
\begin{align}
\log(g_ag_b)^\textnormal{trees}(R)=\frac{1}{2}\sum_{p=1}^2\int_{-\infty}^{+\infty}
\frac{du}{2\pi}\Theta_{pab}(u)\log(1+Y_{p}(u)),\label{tree-su(3)}
\end{align}
where $Y_p$ for $p=1,2$ are solutions of the TBA equations \eqref{su(3)-TBA}. 

Now comes the loop contribution
\begin{align}
\log(g_ag_b)^\textnormal{loops}(R)=\sum_{n\geq 1}\mathcal{C}^\pm_n,
\end{align}
where $\mathcal{C}^\pm_n$ denotes the sum over bosonic/fermionic loops
of length $n$.  Each of these $n$ vertices can be either of type 1 or
2.  The trees growing out of each vertex can be summed to the
Fermi-Dirac factor of each type, by virtue of the relation
\eqref{su(3)-relation}
\begin{align}
\sum_{r}r^2Y_{pr}(u)=\frac{Y_p(u)}{1+Y_p(u)}=f_p(u),\quad p=1,2.
\end{align}
The loop contribution can then be written as a sum over cyclic sets
$\textbf{p}$ of $\mathfrak{C}(\lbrace 1,2\rbrace ^n)$
\begin{align}
\mathcal{C}^\pm_n=\pm\sum_{p_1,...,p_n\in\mathfrak{C}(\lbrace 1,2\rbrace ^n)}
\frac{1}{S(\textbf{p})}\prod_{j=1}^n\int_0^{+\infty}\frac{du_j}{2\pi}f_{p_j}(u_j)
K^\pm_{p_2p_1}(u_2,u_1)....K^\pm_{p_1p_n}(u_1,u_n),
\end{align}
where $S(\textbf{p})$ is the symmetry factor of $\textbf{p}$.  This
sum is nothing but the trace of $2\times 2$ matrices $\hat{K}^\pm$
with elements
\begin{align*}
K^\pm_{pq}(F)(u)=\int_0^{+\infty}\frac{dv}{2\pi}K^\pm_{pq}(u,v)f_q(v)F(v),
\quad p,q\in\lbrace 1,2\rbrace.
\end{align*}
We obtain two Fredholm determinants with $2\times 2$  matrix kernels as a
generalization of \eqref{free-energy-cycles-2}
\begin{align}
\log(g_ag_b)^\textnormal{loops}(R)=\log\det\frac{1-\hat{K}^-}{1-\hat{K}^+}.\label{g-su(3)}
\end{align}
\subsection{g-function of theories with non-diagonal bulk scattering}
A periodic theory with non-diagonal bulk scattering can be diagonalized with the Nested Bethe Ansatz technique. The Bethe equations then involve additional particles with vanishing momentum and energy. These particles called magnons can be regarded as excitations on a spin chain where the physical rapidities are non-dynamical impurities. In particular the number of magnons cannot exceed the number of physical particles. This constraint presents a major obstruction in the implementation of our approach for non-diagonal theories as we cannot  carry out the first step of summing over mode numbers.

When the number of physical particles is large however, magnons can form strings of evenly distributed complex rapidities. As a consequent, the TBA equations are effectively the same as those of a diagonal theories with an infinite number of particles in the spectrum. This means that if we include magnon strings into our cluster expansion and remove the constraint of their numbers, then according to the above analysis we will recover the correct TBA equations.

If we follow this line of logic to non-diagonal open systems, we come to the conclusion that the g-function of these theories is an infinite-dimensional extension of \eqref{tree-su(3)} and \eqref{g-su(3)}. Such direct extension proves however to be problematic. Indeed, as magnons have vanishing energy, there is no driving term in the TBA equations determining their pseudo-energies. As a consequent, the corresponding Y-functions do not vanish at zero temperature limit and so does the g-function. This value of g-function at zero temperature comes from graphs made entirely of magnons. Their contribution is present at any temperature and we propose to get rid of it by  normalizing the finite temperature g-function by its zero temperature limit value 
\begin{align*}
g(R)\to g^{\text{ren}}(R)\equiv \frac{g(R)}{g(R=+\infty)}.
\end{align*}
Under the presence of an infinite tower of magnon strings this normalization can be subtle. We consider a concrete model with string solutions in another paper \cite{Vu2019}.

\section*{Conclusion and outlook}

We propose a graph theory-based method to compute the $g$-function of a
theory with diagonal bulk scattering and diagonal reflection matrices.
The idea is to apply the matrix-tree theorem to write the Jacobians in
the cluster expansion of the partition function by a sum over graphs.
The $g$-function is then written as a sum over trees and loops.  The sum
over trees gives TBA saddle point result while the sum over loops
constitute the two Fredholm determinants.  The method was
generalized to theories of more than one particle type with diagonal bulk scattering and diagonal reflection matrices. We also propose a protocol to obtain g-function of non-diagonal theories solved by Nested Bethe Ansatz.

We would like to point out the relationship between the expression of the g-function and the overlap between an initial state and the ground-state  \eqref{overlap-ground}. The 
normalized overlaps play an important role in the study of out of equilibrium dynamics \cite{1742-5468-2014-6-P06011,1751-8121-47-34-345003,1751-8121-47-14-145003,1742-5468-2014-5-P05006,1742-5468-2018-5-053103} and one point function in AdS/CFT \cite{deLeeuw:2015hxa},\cite{Buhl-Mortensen:2015gfd,deLeeuw:2016umh,deLeeuw:2018mkd}. A direct comparison of the two types of results on the overlaps is not straightforward since they imply different regimes of parameters, but it is an interesting open problem to understand 
the link between the two. 

Several other directions can be investigated in near future. First, we would like to find the missing contribution in our proposition of the excited state g-function. As explained in the main text, our approach yields the correct equations that determine the excited states so it could potentially be modified to produce the corresponding g-function. Second, one can consider the case of non-diagonal reflection matrices.
It would be ideal to have a candidate theory which is sufficiently
simple to be the working example. Last but not least, our method could also be applied in the
hexagon proposal for three point functions in $\mathcal{N}=4$ super
Yang-Mills \cite{Basso:2015zoa},\cite{Basso:2015eqa}. This non-perturbative approach is plagued with divergence when one glues two hexagon form
factors together \cite{Basso:2017muf}. The divergence takes the form of a free energy of particles in the mirror channel. The regularization prescription that leads to this free energy also predicts a finite contribution which bears some similarities to the g-function.

\section*{Acknowledgements}

The authors thank Benjamin Basso for critical comments on nested g-function. We thank Zolt\'an Bajnok for important remarks during the development of this paper, notably for a clear explanation of the different normalizations of the partition function. 
I.K and D.S. would like to thank   
Jean-S\'ebastien Caux, Patrick Dorey, Bal\'azs Pozsgay, Shota Komatsu
 for  discussions, 
and  Yunfeng Jiang, Shota Komatsu, Amit Sever and Edoardo Veskovi   for
sharing the information about an unpublished result concerning the
equivalent of the g-function in the nested case \cite{Jiang:2019xdz}.

 \appendix

\section{A combinatorial proof for the matrix-tree theorem}
\label{appendix:A}

In this appendix we give a direct proof of the matrix-tree theorem in
the form presented in section \ref{matrix-tree}

The aim is to compute the determinant of a $n\times n$ matrix $M$ with
elements
 \begin{align}
 M_{ij}=\(D_i+\sum_{k\neq i}K^+_{ik}\)\,\delta_{ij}-K^-_{ij}\,
 (1-\delta_{ij})\;.\label{M-matrix}
 \end{align}
 in terms of trees and loops made by the elements $K^+_{ij}$ and $K^-_{ij}$.
 
 Compared to the Gaudin matrix \eqref{Laplacian-new}, the notations
 are related as follows
 \begin{align*}
 D_i&\to r_iD_{ab}(u_i)+r_i^2(K_{ii}^+-K_{ii}^-),\\
 K^\pm_{ij}&\to r_ir_jK^\pm_{ij}
 \end{align*}
 
 The tree-matrix theorem states that the determinant of
 \eqref{M-matrix} can be written as a sum over spanning forests for
 the complete graph formed by the $n$ vertices.  The disconnected
 trees contain each either a single loop formed by $K^{-}$ elements,
 or a loop formed by $K^{+}$ elements, or a root associated with
 $D_i$'s.  In this section we do not distinguish between tadpoles and
 roots.  Each $K^{-}$ loop comes with a minus sign.

 To proceed, we express the determinant as a sum over permutations
  \begin{align}
 \det M=\sum_{\sigma\in S_n} (-1)^{s(\sigma)} M_{1\sigma(1)}\ldots M_{n\sigma(n)}\;.
 \end{align}
Each permutation can be decomposed as a product of disjoint cycles of
lengths $k_1,\ldots, k_m$ with $k_1+\ldots+k_m=n$.  Each cycle of
length $k$ comes with a sign $(-1)^{k-1}$, since it involves at least
$k-1$ transpositions.  The structure of the diagonal and
off-diagonal elements of the matrix $M$ is different, one should
consider separately the non-trivial cycles, of length greater than
one, and the trivial ones.  Each non-trivial cycle in the permutation
$\sigma$ gives as a factor a loop formed out of elements $K^-_{ij}$.
For example the cycle $(123)$ will give a contribution
    \begin{align}
    \label{Kmcycle}
(123)\quad \longrightarrow \quad - K^-_{(123)} \equiv -
K^-_{12}\,K^-_{23}\,K^-_{31}\;.
 \end{align}
 The overall minus sign comes from the signature of the cycle times
 $(-1)^k$ form the individual contributions of the matrix elements.
 To discuss the contribution of the trivial cycles, {\it i.e.} of the
 diagonal elements $M_{ii}$, it is convenient to introduce an
 orientation for the elements $K^+_{ij}$, with an arrow going from $j$
 to $i$ (the same can be done for the elements $K^-_{ij}$, so the
 cycle in \eqref{Kmcycle} has an arrow circulating around the loop).
 Let us now consider the factors which contain the diagonal elements
 $M_{ii}$.  For simplicity we are going to consider indices
 $i=1,\ldots, l$, the other cases will be obtained by permutation of
 the indices.  We have
   \begin{align}
    \label{Mii}
M_l\equiv \prod_{i=1}^l M_{ii}=\prod_{i=1}^l \(D_i+\sum_{k\neq i}^l
K^+_{ik}+\sum_{k=l+1}^nK^+_{ik}\)\;,
 \end{align}
 while the complement is given by 
   \begin{align}
    \label{tMii}
\tilde M_l=\sum_r\sum_{{\rm cycles}\in S_{n-l}} (-1)^{r}K^-_{{\rm
cycle}\ 1}\ldots K^-_{{\rm cycle}\ r}\;,
 \end{align}
 where the sum is over the non-trivial cycles involving indices from
 $l+1$ to $n$ and $r$ is the number of cycles.  In \eqref{Mii} we have
 separated in the sums the terms which have both indices in the
 ensemble $\{1,\ldots,l\}$ and those which have one index inside and
 one index outside the ensemble.  The sum in \eqref{Mii} can be
 expanded then as
  \begin{align}
    \label{Miisum}
M_l=\sum_{\alpha_1\cup\alpha_2\cup\alpha_3=\{1,\ldots,l\}}\
\prod_{i\in \alpha_1}\( \sum_{k\neq i}^l K^+_{ik}\)\ \ \prod_{i\in
\alpha_2} D_i \ \ \prod_{i\in \alpha_3} \(\sum_{k=l+1}^nK^+_{ik}\)\;.
 \end{align}
 The terms from the last factor will grow branches attached to the
 loops $K^-_{(s_1\ s_2\ \ldots \ s_m)}$ with indices
 $\{s_1,s_2,\ldots,s_m\}\subset \{l+1,\ldots,n\}$.  \footnote{ A branch is
 associated with a factor of type $K^+_{ij}$, the origin of the branch
 being the second index, $j$ and the tips to the first index $i$.} The
 tips of these branches belong to the ensemble $\alpha_3$.  The second
 factor in \eqref{Miisum} give roots in the ensemble $\alpha_2$.
 
 The first factor $\prod_{i\in \alpha_1} \sum_{k\neq i}^l K^+_{ik}$
 has a more complicated structure.  In the case when
 $\alpha_1=\{1,\ldots,l\}$, it contains at least one loop of type
 $K^+_{(s_1\ s_2\ \ldots \ s_m)}$ with indices in
 $\{s_1,s_2,\ldots,s_m\}\subset \{1,\ldots,l\}$.  The reason is that each
 term in the sum has the structure
  \begin{align}
    \label{Kpcycle}
 K^+_{1\star}\,K^+_{2\star}\,\ldots K^+_{l\star}\;,
 \end{align}
 where $\star$ denotes an arbitrary second index not equal to the
 first one.  Let us suppose that one of the indices denoted by a star
 is the beginning of a tree.  Because the same index appears as a
 first index as well, we conclude that the corresponding vertex is
 also the tip of a branch, so it belongs to a loop.  In a single
 factor of the type \eqref{Kpcycle} there can be several loops, and
 multiple branches can grow out from these loops.  Two different loops
 cannot be joined by a branch, because in this case two branches would
 join at their tips, and this is forbidden by the structure in
 \eqref{Kpcycle} where each tip of a branch is different from the
 others.  We conclude that when $\alpha_1=\{1,\ldots,l\}$ the
 corresponding contribution is that of disjoint graphs with a single
 loop each and with branching growing out of them, spanning the
 ensemble of vertices $\{1,\ldots,l\}$.
 
  When $\alpha_1\neq \{1,\ldots,l\}$ one should repeat again the
  procedure of splitting the sum over indices,
    \begin{align}
    \label{sumsplit}
\prod_{i\in \alpha_1}\( \sum_{k\neq i}^l K^+_{ik}\)&= \prod_{i\in
\alpha_1} \(\sum_{k\neq i; k\in \alpha_1}K^+_{ik}+\sum_{ k\notin
\alpha_1}K^+_{ik}\)\\ \no &=
\sum_{\alpha_{11}\cup\alpha_{12}=\alpha_1} \prod_{i\in
\alpha_{11}}\(\sum_{k\neq i; k\in \alpha_1}K^+_{ik} \)\prod_{i\in
\alpha_{12}}\(\sum_{ k\in \alpha_2\cup\alpha_3}K^+_{ik}\)\;.
 \end{align}
 The terms from the second product in the second line above will add a
 new layer of branches from the branches already grown from the loops
 of type $K^-_{(s_1\ s_2\ \ldots \ s_m)}$, if $k\in \alpha_3$, or will
 grow branches from the roots $D_i$, if $k\in \alpha_2$.  The new
 branches have tips in the ensemble $\alpha_{12}$.  The terms in the
 first product will be treated as in the previous stage.  The
 procedure will be repeated until all the indices are exhausted.
 
 We conclude that after repeating the procedure we are left with an
 ensemble of disconnected (generalised) trees each growing out from
 \begin{itemize}
 \item
 a loop of type $K^-_{(s_1\ s_2\ \ldots \ s_m)}$ or
 \item
  a loop of type $K^+_{(s_1\ s_2\ \ldots \ s_m)}$ or
  \item
  a root of type $D_i$ 
 \end{itemize}
 spanning the indices $\{1,\ldots,n\}$.

 \section{A field-theoretical proof of the matrix-tree theorem}

      \def\thth{ { \bm {\theta } }}
      
  To begin with, we write the matrix $M$ defined by \re{M-matrix} in 
  a slightly different form,
   \be
   \la{M-matrixB}
 M_{ij} = \hat M_i \, \d_{ij} - K^-_{ij},
 \qquad
\hat  M_i \equiv \hat D_i + \sum_{k=1}^n  K_{ik}^+.
 \ee
Note that in this writing the second term does not vanish on the diagonal.  
   Compared to the Gaudin matrix \eqref{Laplacian-new}, the notations here 
 are related as follows
 \begin{align*}
\hat  D_i&\to r_iD_{ab}(u_i) ,\\
 K^\pm_{ij}&\to r_ir_jK^\pm_{ij}.
 \end{align*}

The starting point is the representation of the determinant
(\ref{M-matrix}) as an integral with respect to $n$ pairs of
grassmannian variables $\th_i, \bar\th_i \ (i=1,..., n)$.
 The determinant of any matrix $M= \{M_{jk}\}_{j,k =1}^m$ can be
 written as an integral over $n$ pairs of grassmannian variables
 $\thth= \{ \th_1,...,\th_m\}$ and $\bar \thth= \{ \bar \th_1,...,\bar
 \th_m\} ^{T}$:
  \be\begin{aligned} \la{fermirepM} \det M= \int \prod _{i=1}^n d
  \th_i d\bar\th_i\ e^{ \sum_{ij} \bar\th_i M_{ij} \th_j}.
\end{aligned}
\ee
 For a matrix of the type  \re{M-matrixB} we want to express the determinant in terms of the quantities $\hat D_i$ and $K^\pm_{ij}$. 
 For that we first  expand the exponential of the diagonal part using the
 nilpotent property of the grassmannian variables,
\be\begin{aligned}
\la{fermirepMd}
 \det M &
 &=(-1)^n \int \prod _{j=1}^n d\theta_jd\bar \theta_j\ (1+
 \bar\theta_j\theta_j \hat M_j) \, e^{- \sum _{j,k=1}^n \bar \theta_j
 K^-_{jk} \theta_k}
.
\end{aligned} 
 \ee
   Now we go to the dual variables $\bar\psi_i, \psi_i$, related to
   the original ones by a Hubbard-Stratonovich transformation
\be\begin{aligned}
 \det M
&=
\int \prod _{j=1}^n d\theta_jd\bar\theta_jd\psi_jd\bar\psi_j\
e^{- \sum _{j,k} \bar  \theta_{j}   K ^-_{jk} \theta_k
-\sum_j (\bar\theta_j\psi_j + \theta_j\bar \psi_j) }
\prod_j \(  \bar\psi_j\psi_j+\hat M_j\).
\end{aligned}\ee
Here we used the obvious identities for grassmanian integration
\be
\int d\psi d\bar\psi \ 
e^{\bar\th\psi + \th \bar\psi}
= \bar\th \th\, ,
\quad
\int d\psi d\bar\psi \ 
e^{\bar\th\psi + \th \bar\psi}
 \bar\psi  \psi = 1.
\ee
This gaussian  integral is evaluated  by performing all   Wick 
contractions $\<\bar  \psi_j\psi_k\> = K^{-}_{jk}$.
 Symbolically
 \be\begin{aligned} \la{Mferm} \det M &= \Big\langle\prod_{j=1}^m
 \Big( \bar\psi_j \psi_j+\hat M_j \Big)\Big\rangle_{\text{Wick}}, \quad \,
 \<\bar \psi_j\psi_k\> = K^{-}_{jk}\, .  \end{aligned}\ee

In a similar way, we will  introduce the 
piece   $\sum_k K^+_{jk}$ in  $M_j$  through the expectation value 
with respect to $n$   pairs of bosonic variables
 $\vp_i,\bar\vp_i \ (i=1, ..., n)$
   \be\begin{aligned} \prod_{j=1}^n (\bar\psi_i\psi_i + \hat M_j)&= e^{
   {\sum_{j,k=1}^n {\p\over \p \vp_j}K^+_{jk}{\p\over \p \bar \vp_k}}}
   \ \prod_{j=1}^n e^{\vp_j}\[ \hat D_j +\bar\vp_j+ \bar\psi_i \psi_i \]
   \Bigg|_{\vp_j= \bar\vp_j=0}.  \end{aligned}\ee
  Equivalently one can represent the rhs as an expectation value with
  respect to $n$ pairs of quantum bosonic variables with correlator
  $\< \bar\vp_i\vp_j\> = K^+_{ij}$, with all other correlators
  vanishing.  Together with \re{Mferm}, this yields the following
  representation of the determinant as an expectation value
\be\begin{aligned} \det M &= \left\langle \prod_{j=1}^m ( \hat D_j +
\bar\varphi _j + \bar\psi_j\psi_j )\, e^{\varphi _j}
\right\rangle_{\mathrm{Wick}}, \la{QFTrep}\end{aligned} \ee
with the non-zero bosonic and   fermionic  propagators given respectively by
\be\begin{aligned} \langle \varphi _j\bar\varphi _k\rangle = K^+_{jk},
\quad \langle \psi_j\bar\psi_k\rangle = K^-_{jk}.
\la{QFTprops}\end{aligned} \ee
 Performing all possible fermionic and bosonic Wick contractions
 generates the forest expansion of the determinant.  The expectation
 value is a sum of all Feynman graphs (in general disconnected) whose
 vertices cover the set $\{1, 2, ..., n\}$ once and only once.  Each
 Feynman graph consists of vertices connected by propagators.  The
 correlator $\< \bar \vp_i \vp_j\> = K^+_{ji}$ is represented by an
 oriented line pointing from $i$ to $j$.  The correlator $\langle
 \bar\psi_i\psi_j\rangle = K_{ij}^-$ is represented by an oriented
 dotted line.  At each vertex there is at most one incoming line while
 the number of the outgoing lines is unrestricted.  The vertices with
 one incoming line have weight 1 while the vertices with only outgoing
 lines have weight $\hat D_i$.  If a vertex has a fermionic incoming line,
 then it must have one fermionic outgoing lines and an unrestricted
 number of outgoing bosonic lines.  There is only one such vertex per
 connected tree and it corresponds to the root.  Each connected graph
 can have at most one loop, fermionic or bosonic.  The fermionic loops
 have extra factor $(-1)$.


\begin{thebibliography}{10}

\bibitem{Ghoshal:1993tm}
Subir Ghoshal and Alexander~B. Zamolodchikov.
\newblock {Boundary S matrix and boundary state in two-dimensional integrable
  quantum field theory}.
\newblock {\em Int. J. Mod. Phys.}, A9:3841--3886, 1994.
\newblock [Erratum: Int. J. Mod. Phys.A9,4353(1994)].

\bibitem{Affleck:1991tk}
Ian Affleck and Andreas W.~W. Ludwig.
\newblock {Universal noninteger 'ground state degeneracy' in critical quantum
  systems}.
\newblock {\em Phys. Rev. Lett.}, 67:161--164, 1991.

\bibitem{LeClair:1995uf}
A.~LeClair, G.~Mussardo, H.~Saleur, and S.~Skorik.
\newblock {Boundary energy and boundary states in integrable quantum field
  theories}.
\newblock {\em Nucl. Phys.}, B453:581--618, 1995.

\bibitem{Woynarovich:2004gc}
F.~Woynarovich.
\newblock {O(1) contribution of saddle point fluctuations to the free energy of
  Bethe Ansatz systems}.
\newblock {\em Nucl. Phys.}, B700:331--360, 2004.

\bibitem{Dorey:2004xk}
Patrick Dorey, Davide Fioravanti, Chaiho Rim, and Roberto Tateo.
\newblock {Integrable quantum field theory with boundaries: The Exact g
  function}.
\newblock {\em Nucl. Phys.}, B696:445--467, 2004.

\bibitem{Pozsgay:2010tv}
Balazs Pozsgay.
\newblock {On O(1) contributions to the free energy in Bethe Ansatz systems:
  The Exact g-function}.
\newblock {\em JHEP}, 08:090, 2010.

\bibitem{Woynarovich:2010wt}
F.~Woynarovich.
\newblock {On the normalization of the partition function of Bethe Ansatz
  systems}.
\newblock {\em Nucl. Phys.}, B852:269--286, 2011.

\bibitem{Chaiken82acombinatorial}
Seth Chaiken.
\newblock A combinatorial proof of the all minors matrix tree theorem, 1982.

\bibitem{Kostov:2018ckg}
Ivan Kostov, Didina Serban, and Dinh-Long Vu.
\newblock {TBA and tree expansion}.
\newblock In {\em {12th International Workshop on Lie Theory and Its
  Applications in Physics (LT-12) Varna, Bulgaria, June 19-25, 2017}}, 2018.

\bibitem{article}
Go~Kato and Miki Wadati.
\newblock Graphical representation of the partition function of a
  one-dimensional delta-function bose gas.
\newblock 42:4883--4893, 10 2001.

\bibitem{PhysRevE.63.036106}
Go~Kato and Miki Wadati.
\newblock Partition function for a one-dimensional \ensuremath{\delta}-function
  bose gas.
\newblock {\em Phys. Rev. E}, 63:036106, Feb 2001.

\bibitem{10.21468/SciPostPhys.6.2.023}
Dinh-Long Vu and Takato Yoshimura.
\newblock {Equations of state in generalized hydrodynamics}.
\newblock {\em SciPost Phys.}, 6:23, 2019.

\bibitem{Vu2019}
Dinh-Long Vu, Ivan Kostov, and Didina Serban.
\newblock Boundary entropy of integrable perturbed su (2)k wznw.
\newblock {\em Journal of High Energy Physics}, 2019(8):154, Aug 2019.

\bibitem{Kormos:2010ae}
Marton Kormos and Balazs Pozsgay.
\newblock {One-Point Functions in Massive Integrable QFT with Boundaries}.
\newblock {\em JHEP}, 04:112, 2010.

\bibitem{1742-5468-2014-6-P06011}
Bal\'azs Pozsgay.
\newblock Overlaps between eigenstates of the xxz spin-1/2 chain and a class of
  simple product states.
\newblock {\em Journal of Statistical Mechanics: Theory and Experiment},
  2014(6):P06011, 2014.

\bibitem{1751-8121-47-34-345003}
M~Brockmann, J~De Nardis, B~Wouters, and J-S Caux.
\newblock N\'eel-xxz state overlaps: odd particle numbers and lieb-liniger
  scaling limit.
\newblock {\em Journal of Physics A: Mathematical and Theoretical},
  47(34):345003, 2014.

\bibitem{1751-8121-47-14-145003}
M~Brockmann, J~De Nardis, B~Wouters, and J-S Caux.
\newblock A gaudin-like determinant for overlaps of n\'eel and xxz bethe
  states.
\newblock {\em Journal of Physics A: Mathematical and Theoretical},
  47(14):145003, 2014.

\bibitem{1742-5468-2014-5-P05006}
M~Brockmann.
\newblock Overlaps of q-raised n\'eel states with xxz bethe states and their
  relation to the lieb-liniger bose gas.
\newblock {\em Journal of Statistical Mechanics: Theory and Experiment},
  2014(5):P05006, 2014.

\bibitem{1742-5468-2018-5-053103}
B~Pozsgay.
\newblock Overlaps with arbitrary two-site states in the xxz spin chain.
\newblock {\em Journal of Statistical Mechanics: Theory and Experiment},
  2018(5):053103, 2018.

\bibitem{deLeeuw:2015hxa}
Marius de~Leeuw, Charlotte Kristjansen, and Konstantin Zarembo.
\newblock {One-point Functions in Defect CFT and Integrability}.
\newblock {\em JHEP}, 08:098, 2015.

\bibitem{Buhl-Mortensen:2015gfd}
Isak Buhl-Mortensen, Marius de~Leeuw, Charlotte Kristjansen, and Konstantin
  Zarembo.
\newblock {One-point Functions in AdS/dCFT from Matrix Product States}.
\newblock {\em JHEP}, 02:052, 2016.

\bibitem{deLeeuw:2016umh}
Marius de~Leeuw, Charlotte Kristjansen, and Stefano Mori.
\newblock {AdS/dCFT one-point functions of the SU(3) sector}.
\newblock {\em Phys. Lett.}, B763:197--202, 2016.

\bibitem{deLeeuw:2018mkd}
Marius De~Leeuw, Charlotte Kristjansen, and Georgios Linardopoulos.
\newblock {Scalar one-point functions and matrix product states of AdS/dCFT}.
\newblock {\em Phys. Lett.}, B781:238--243, 2018.

\bibitem{Basso:2015zoa}
Benjamin Basso, Shota Komatsu, and Pedro Vieira.
\newblock {Structure Constants and Integrable Bootstrap in Planar N=4 SYM
  Theory}.
\newblock 2015.

\bibitem{Basso:2015eqa}
Benjamin Basso, Vasco Goncalves, Shota Komatsu, and Pedro Vieira.
\newblock {Gluing Hexagons at Three Loops}.
\newblock {\em Nucl. Phys.}, B907:695--716, 2016.

\bibitem{Basso:2017muf}
Benjamin Basso, Vasco Goncalves, and Shota Komatsu.
\newblock {Structure constants at wrapping order}.
\newblock {\em JHEP}, 05:124, 2017.

\bibitem{Jiang:2019xdz}
Y.~Jiang, S.~Komatsu, and Edoardo Vescovi.
\newblock {Structure Constants in $\mathcal{N}=4$ SYM at Finite Coupling as
  Worldsheet $g$-Function}.
\newblock 2019.

\end{thebibliography}
\end{document}